\documentclass[12pt]{article}

\usepackage{amsmath,amsfonts,latexsym,fullpage,graphicx,vmargin,natbib}
\usepackage{amssymb, mathrsfs}
\usepackage{pstricks}
\usepackage{subcaption}
\usepackage{amsmath}
\usepackage{multirow}
\usepackage{comment}
\usepackage[inline]{enumitem}
\usepackage{hyperref}
\usepackage[normalem]{ulem}
\usepackage{placeins}
\newcommand{\iid}{\stackrel{\mbox{\scriptsize iid}}{\sim}}
\newcommand{\ind}{\stackrel{\mbox{\scriptsize ind}}{\sim}}
\newcommand{\bm}[1]{\mbox{\boldmath{$#1$}}}

\newcommand{\calN}{\mathcal{N}}

\newcommand{\mtilde}{\widetilde{m}}
\newcommand{\Law}{\mathcal L}

\DeclareMathOperator{\Cov}{Cov}

\newcommand{\alr}{\text{alr}}
\newcommand{\wtilde}{\widetilde{w}}
\DeclareRobustCommand{\rchi}{{\mathpalette\irchi\relax}}
\newcommand{\irchi}[2]{\raisebox{\depth}{$#1\chi$}}

\newtheorem{prop}{Proposition}

\newcommand{\R}{\mathbb{R}}

\newcommand{\Y}{\mathbb{Y}}
\newcommand{\E}{\mathbb{E}}

\newcommand{\e}{\mathrm{e}}

\usepackage{ulem}

\newcommand{\virgolette}[1]{``#1''}

\begin{document}

\title{\bf Spatially dependent mixture models via the Logistic Multivariate CAR prior}
\author{
  Mario Beraha\thanks{Department of Mathematics, Politecnico di Milano}
  \thanks{Department of Computer Science, Universit\`a di Bologna},
  Matteo Pegoraro\thanks{MOX - Department of Mathematics, Politecnico di Milano},
  Riccardo Peli\footnotemark[3] and
  Alessandra Guglielmi\footnotemark[1]
}
\date{08 June 2021}

\maketitle

%\author[DIP,BOL]{Mario Beraha}
%\ead{mario.beraha@polimi.it}
%
%\author[MOX]{Matteo Pegoraro}
%\ead{matteo.pegoraro@polimi.it}
%
%\author[MOX]{Riccardo Peli}
%\ead{riccardo.peli@polimi.it}
%
%\author[DIP]{Alessandra Guglielmi}
%\ead{alessandra.guglielmi@polimi.it}

%\affiliation[DIP]{organization={Department of Mathematics, Politecnico di Milano},
%	addressline={},
%	city={},
%	postcode={},
%	country={}}
%
%\affiliation[BOL]{organization={Department of Computer Science, Universit\`a di Bologna},
%	addressline={},
%	city={},
%	postcode={},
%	country={}}
%
%\affiliation[MOX]{organization={MOX, Department of Mathematics, Politecnico di Milano},
%	addressline={},
%	city={},
%	postcode={},
%	country={}}

\begin{abstract}
We consider the problem of spatially dependent areal data, where for each
area independent observations are available, and propose to model
the density of each area through a finite mixture of Gaussian distributions.
The spatial dependence is introduced via a novel joint distribution for
a collection of vectors in the simplex, that we term logisticMCAR.
We show that salient features of the logisticMCAR distribution  
can be described analytically, and that a suitable augmentation scheme based on the 
P{\'o}lya-Gamma identity allows to derive an efficient Markov Chain Monte Carlo
algorithm.
When compared to competitors, our model has proved to better estimate densities in different (disconnected) areal locations when they have different characteristics.
We discuss an application on a real dataset of Airbnb listings in the city
of Amsterdam, also showing how to easily incorporate for additional covariate
information in the model.
\end{abstract}


\section{Introduction}
In spatial statistics, it is often assumed that data in neighboring locations are likely to behave more similarly than those that are far away.
Thus, inference and prediction methods have been developed to take into account spatial dependence. 
Spatial data are classified into three main categories, according to \cite{cressie1992statistics}:  geostatistical data, for which an exact location is known for each observation, areal (or lattice) data, when each observation is associated to a specific area or node in a lattice, and point patterns, where the object of the inference is the event location.
Examples of the first are environmental applications \citep[see][]{enviromentalgeostat} and geological reservoir characterization for oil and gas recovery \citep[see][for examples]{deutsch2002geostatistical}. A recent review paper on 
statistical models for areal data is \cite{banerjee2016spatial}, which focuses on disease mapping and spatial survival analysis. Point patterns are often employed in ecology, as described in \cite{Velazquez2016}.
See also the textbook  by \cite{banerjee2014hierarchical} for data classification, applications and statistical models and techniques for spatially dependent data.

\subsection{Setup}
We focus on areal data, and, in particular,
we consider the problem of modeling data 
%\pego{data generating densities?}} 
from $I$ different groups, where each group corresponds to a specific areal location.
More in detail, we assume that the spatial domain $\Omega$ is divided into $I$ areas and, for each
area, there is a vector of observations $\bm y_i = (y_{i1}, \ldots, y_{iN_i})$ from the same variable, each value $y_{ij}$ corresponding to a different subject $j$ in area $i$.
The goal of this manuscript is the proposal of a statistical model, for data $\{\bm y_i, i=1,\ldots,I\}$,  accounting for
dependence arising from spatial proximity while being flexible enough to model data 
that do not fit standard parametric distributions. 
We further assume that data, within each areal unit $i$, are independent and identically distributed (i.i.d.) from an area-specific density $f_i$;
the problem we address is the joint estimation of spatially dependent densities $f_1,\ldots,f_I$.
We take the Bayesian viewpoint and we specify a prior for dependent densities $(f_1,\ldots,f_I)$ that encourages distributions associated to areas that are spatially close
to be more similar than those associated to areas that are far away. 
Relaxing the assumption of identically distributed  observations within each area is straightforward in the regression context, i.e. when covariates for each subject are available.  

As motivating application, we consider publicly available data on Airbnb listings in the city of Amsterdam (NL). 
Airbnb is the largest vacation rental marketplace. In recent years it has 
been debated that Airbnb has deeply transformed the social structure
of major touristic cities, as Amsterdam \citep{van2016airbnb}, Barcelona \citep{garcia2018urban} and several US cities \citep{wachsmuth2018airbnb},  driving up property prices and disrupting communities.
The application dataset consists of more than $17,000$ listings spread over neighborhoods in Amsterdam. 
Our goal is to predict the nightly price of a new listing, with information given by covariates,
taking into account the spatial dependence. Such a model can be of interest to a \virgolette{new}  lessor wishing to rent their house or flat on Airbnb. The area-specific estimate of the density might allow the lessor to understand the full market of renting apartments in his/her neighborhood, unlike a
simple point estimate of the average price.
The lessor might also understand if it is worth making home improvements in order to get a higher rent or assessing, for instance,
the posterior predictive probability of the rent being above some threshold. 

 A peculiar feature of the municipality of Amsterdam is that three neighborhoods are  not connected to the rest of the city but among themselves (see, for instance, Figure~\ref{fig:nlist_mean_std_by_neigh}), i.e. there are two different connected components in the adjacency graph of neighborhoods.
It is likely that the nightly prices exhibit substantially different behavior when comparing one component to the other. Hence, we want to build a model that encourages sharing of information across neighboring areas, but does not force densities belonging to different components to be similar a priori.

Compared to more traditional spatial regression techniques such as eigenvector spatial filtering \citep[see][for a review]{griffith2019spatial}, geographically weighted regression \citep{brunsdon1998geographically} or the models in the \texttt{R} package \textit{CARBayes} \citep{carbayes}, 
our approach does not make distributional assumptions (such as assuming Gaussian-distributed responses) and our focus here is on density modeling and estimation and density regression via mixture models.

\subsection{Previous work on Bayesian spatial density modeling}

To model our distributions we resort to the well established class of mixture models \citep{fruhwirth2019handbook}, that are a classical tool for density estimation.
In the Bayesian nonparametric setting, since \cite{maceachern2000dependent},
a great effort has been dedicated to modeling a set of related, though not identical, distributions.
Dealing with spatial processes, \cite{gelfand2005bayesian} and \cite{duan2007generalized}
developed a spatial dependent Dirichlet process as random-effects distribution in the context of point-reference data. The stick-breaking representation of the Dirichlet process allows all the models built from it to be considered as infinite mixture models. 
 Starting from the stick-breaking representation of the dependent Dirichlet process in the particular case of \emph{single atoms} (atoms not indexed by covariates), \cite{dunson2008kernel} proposed the kernel stick-breaking process mixtures; 
spatial extensions of these type of mixtures have been developed
to accommodate for general covariates and spatial locations for geostatistical data, such as, e.g., \cite{rodriguez2011nonparametric}  and \cite{ren2011logistic}.
\cite{jo2017dependent} considered mixture models based on species sampling priors where the spatial dependence is introduced through a Gaussian multivariate conditional autoregressive \citep[CAR,][]{besag1974spatial} model on a suitable transformation of the weights. Despite their focus being on point-referenced data, their model can be easily extended to areal data, as we do in Sections~\ref{sec:comparison} and \ref{sec:simulation} for a comparison with our approach.
The idea of building spatial dependence in mixture models through a CAR distribution on latent variables is also shared by \cite{li2015bayesian},
where the authors propose an area-dependent Dirichlet process that can also formally identify boundaries between areas, and \cite{zhou2015spatio}, that use the trick of normalization of CAR distributions to time-varying weights in a rather complex application with focus on estimation of ambulance demand.

Despite the theoretical  properties of Bayesian nonparametric mixtures,  computing the posterior inference in this
setting may yield computational issues. In fact, typical MCMC algorithms here would need to marginalize out
the infinite dimensional distribution from the joint distribution of data and parameters, which might not be possible for models
exhibiting a complicate dependence structure such as those mentioned above.  As an alternative, finite-dimensional approximations of the infinite  mixture representation are typically used in the MCMC algorithms.  
However, as recently pointed out by \cite{rigonpitmanyor}, the truncation procedure, for some models, 
might yield unwanted assumptions on the prior distribution of the 
number of clusters. 

%Other authors have considered similar tricks, e.g.
%\cite{jo2017dependent}, who build on the CAR model by  \cite{clayton1987empirical}.

\subsection{Our contribution and outline}
 
In this paper, we consider a finite mixture model, where the number $H$ of components is fixed. Finite mixtures are particularly suited for 
the problem of modeling areal densities because (i) they adapt capturing the spatial dependence more than nonparametric mixtures,  
mainly because the weights of the finite mixtures are not forced
to decrease exponentially fast to $0$ as in many Bayesian nonparametric mixture models, and (ii) posterior inference under 
finite mixtures is extremely simple and admits  efficient parallel code (unlike nonparametric models),  
thus helping our model scaling up as the size of the dataset
increases. See \cite{fruhwirth2006finite} and \cite{celeux2018model} for more insights on finite mixtures.

The first contribution of this work is the introduction of a joint distribution for a collection $\bm w_1, \ldots, \bm w_I$ of $I$
vectors in the simplex $S^H$, reflecting the areal proximity structure in the distribution, through a logistic transformation of Gaussian multivariate CAR models. This distribution has been termed here the logistic MCAR distribution. 
Other authors have considered similar tricks, e.g.
\cite{jo2017dependent}, who build on the CAR model by  \cite{clayton1987empirical}.

A second contribution of this work is the proposal of a finite Gaussian
mixture model for each of the $I$ area-related densities, keeping in mind the flexibility of the Gaussian
mixtures to accurately approximate smooth densities. 
We let all the mixtures
share the same set of atoms, while introducing similarity between the different mixtures through the logistic MCAR distribution, that 
we use as a prior for the weights of the mixtures. 
Through simulated data examples and the Airbnb application we show how 
specific features of the proposed model include (i) a sparse mixture specification as meant in \cite{malsiner2016model}  and (ii) densities corresponding to areal units which belong to two  different connected components in the proximity graph may behave differently.
%(\bale insistere sullo stesso concetto? Dire altro? \eale).
We discuss this last particular point in our data illustrations.
   
A third contribution of this paper  is that we show how the full conditionals
of the mixture weights can be sampled using a Gibbs sampler based on the
P{\'o}lya-Gamma distribution, without resorting to Metropolis-Hastings steps, by exploiting a data augmentation scheme.
As discussed in \cite{polson2013bayesian}, this update can lead to major
improvements in the mixing of the chain. 
Our examples focus on continuous responses and the Gaussian kernel, though extensions to different kernels
can be straightforwardly accommodated in our framework.
 
The rest of this article is organized as follows. 
Section~\ref{sec:prelim} gives background on finite mixture models and the geometry on the finite-dimensional simplex. Section~\ref{sec:logisticMCAR} illustrates the definition and properties of the joint distribution
 of a collection of $I$ 
vectors in the simplex, taking into account the
underlying spatial proximity matrix. Our area-dependent mixture model is illustrated in Section~\ref{sec:themodel}, and the sparse mixture specification is detailed in Section~\ref{sec:prior_mtilde}; Section~\ref{sec:comparison} discusses on the differences
 between our spatial prior and that in \cite{jo2017dependent}.  Section~\ref{sec:gibbs} sketches the Gibbs sampler to compute the posterior  and
Section~\ref{sec:simulation} presents results from two simulation studies with comparison with competitor models. The application to Airbnb Amsterdam is discussed in 
Section~\ref{sec:airbnb}, where we propose two generalizations of our area-dependent mixture model to include subject-specific covariates and relaxing  the identity in distribution assumption within each area.
We conclude in Section~\ref{sec:discussion} with final comments and discussion.
The Appendix collects the proofs for the theoretical
results, Monte Carlo simulations from the joint distribution  of the $I$ vectors in the simplex, full description of the Gibbs sampler, as well as additional plots and tables for the examples. 
Codes of our MCMC algorithm for simulated data and Airbnb Amsterdam application  has been implemented \texttt{C++} and \texttt{Python} and 
is available at \url{https://github.com/mberaha/spatial_mixtures}.

\section{Preliminaries}
\label{sec:prelim}
\subsection{Mixture Models}

For any areal unit $i=1,\ldots,I$ and subject $j=1,\ldots,N_i$, we assume observation  $y_{ij} \in \Y \subset \R^p$.
In this paper, we fix $p=1$, but multivariate responses can be straightforwardly accommodated in our context.
A flexible model for the density in each  area can be constructed by  assuming a finite mixture, specifically
\begin{equation}\label{eq:ex_mixtures}
    y_{ij} \mid \bm w_i, \bm \tau_i \iid f_{i}(\cdot) = \sum_{h=1}^H w_{ih} k(\cdot\mid \tau_{ih}) \quad j=1, \ldots, N_i
\end{equation}
where $k(\cdot \mid \tau)$ is a density on $\Y$ for any $\tau \in \Theta$, and $\Theta$ is the parameter space. 
%In our example we will consider $\Theta = \mathbb{R} \times [0, +\infty)$ but our methodology is not limited to this particular case. 
Each vector $\bm w_i = (w_{i1}, \ldots, w_{iH})^T$, the weights of the mixture \eqref{eq:ex_mixtures}, 
belongs to the $H-1$ dimensional simplex $S^H$, where 
\begin{equation}
S^H:=\{(z_1,\ldots,z_H)\in\R^H: 0\leq z_h\leq 1 , h=1,\ldots,H, \ \sum_{h=1}^H z_h=1\}
\label{eq:def_simplex}
\end{equation}
and $\bm \tau_{i} = (\tau_{i1}, \ldots, \tau_{iH})^T$ are parameters in $\Theta^H$.
In this paper, we refer to $\bm \tau_i$ and $\bm w_{i}$ as the \emph{atoms} and the \textit{weights} of the mixture $f_i$. 

%%\mario{
%%Finite mixtures are a popular tool both in the Bayesian and frequentist
%%context when the statistical goal is density estimation,  clustering of the subjects in the sample or parameter estimation in the context of heterogeneous populations.
%%For a detailed review of mixture models from the Bayesian point of view see
%%\cite{fruhwirth2006finite} and \cite{fruhwirth2019handbook}. 
%%}\footnote{secondo me si puo' tagliare, lo diciamo gia' nell'introduction}
 
Our goal is to introduce dependence between mixtures such  that  data in  neighboring areas are more likely to be modeled with similar distributions than data in far areal units.
A general mixture model like \eqref{eq:ex_mixtures} would require
to model jointly both the atoms and the weights of all the mixtures, in order to obtain
a dependence structure suitable for spatial applications, which can be a challenging task in general, unless we consider a very specific application.
In our approach instead,
borrowing ideas from the single atom dependent Dirichlet processes, 
%%\citep[see][for notation]{barrientos_etal2012},  
we constrain all the
atoms across the different areas to be equal, i.e. $\bm \tau_1 = \bm \tau_2, \ldots = \bm \tau_I = \bm \tau$, and focus only on the weights of the mixtures.
In this way, a sufficient condition for two different mixtures to be similar is 
to have similar weights.
In general, it is more difficult to define mixtures with area-dependent weights than generalizing to area-dependent weights and atoms, since simulation algorithms for models based on standard mixture models can usually be adapted
with few modifications to dependent atoms. 
%\mario{Among the main contributions of this work, there is the construction of a prior for the collection of weights $\bm w_1, \ldots, \bm w_I$ in such a way that weights associated to close areas are more similar than weights associated to areas farther away.}\footnote{\mario{e' la terza volta che lo diciamo... si puo' togliere?}}
 
When the goal of the inference is cluster estimation,
the choice of $H$ might become crucial. 
%%While the choice of a small value for $H$ could not lead to identify all the \virgolette{true groups} in the data, choosing a value too large, if not paired with a suitable
%%prior on the weights, might yield a cluster estimate with many small (and not well separated) groups. 
An alternative consists in assuming $H$ random, including it in the state space of the MCMC algorithm; see, for instance, \cite{nobile1994bayesian}.
However, 
%as discussed in \cite{miller2018mixture} and \cite{argiento2019infinity},
inference in this setting can be computationally intensive 
%(\bale MARIO:  io DIREI computationally expensive \eale), 
as it needs to rely either on specifically designed trans-dimensional MCMC moves 
%%whose design depends on the specific model 
\citep[see][]{green1995reversible,richardson1997bayesian}, or to numerically evaluate infinite series,
as in \cite{miller2018mixture} and in the marginal sampler in \cite{argiento2019infinity}.
On the other hand, sparse mixture models, as meant in
\cite{malsiner2016model},
assume a large value for $H$, larger than needed, and 
a prior %%on the vector weights $\bm w_i \in S^H$ 
assigning large mass to configurations where the weights  of the superfluous components assume
values close to zero. 
This implies  that  the prior number of \emph{non-empty} components 
(i.e. components where at least one observation is allocated to) is significantly smaller than $H$.

In Section~\ref{sec:logisticMCAR} we propose a prior distribution for 
$(\bm w_1, \ldots, \bm w_I)$,  in such a way that weights associated to close areas are more similar than weights associated to areas farther away, by constructing 
a Markov random field for random vectors with bounded sum.
Moreover, by assuming a prior on the hyperparameters, we 
also show that this prior can induce sparsity in the mixture (see Section~\ref{sec:prior_mtilde}) as in \cite{malsiner2016model}.

\subsection{Geometry on the simplex $S^H$}

The simplex $S^H\subset \R^H$ defined in \eqref{eq:def_simplex} is not a vector 
subspace of $\mathbb{R}^H$.
However, $S^H$ is a vector space when equipped with the so-called Aitchison geometry, that defines the operation of \textit{perturbation} (analogous of addition), \textit{powering} (analogous of multiplication by scalar) and \textit{inner product}.
If $\bm w, \bm w_1, \bm w_2 \in S^H$, $\alpha \in \mathbb{R}$   we have
\begin{align*}
     \bm w_1 \oplus \bm w_2 &= \mathcal C(w_{11} w_{21}, \ldots, w_{1H} w_{2H}) := \left( \frac{w_{11} w_{21}}{\sum_{i=1}^H w_{1i} w_{2i}} \dots \frac{w_{1H} w_{2H}}{\sum_{i=1}^H w_{1i} w_{2i}} \right) \\
    \alpha \odot \bm w &= \mathcal{C} (w_1^\alpha, \dots, w_H^\alpha)  \qquad   \qquad %\\
    \langle \bm w_1, \bm w_2 \rangle = \frac{1}{2H} \sum_{i, j = 1}^H \log \frac{w_{1i}}{w_{1j}} \log \frac{w_{2i}}{w_{2j}}
\end{align*}
where $\mathcal{C}$ denotes the \textit{closure}, or normalization (i.e. dividing each element by the sum of all the elements) of a vector in $\R^H$. The symbols $\oplus$, $\odot$ and $  \langle \cdot, \cdot \rangle$ denote perturbation, powering and inner product, respectively.

Many maps from $S^H$ to $\mathbb{R}^{H-1}$ are available in the literature. For our purpose we focus on the bijective additive log-ratio transformation (alr), defined by  $\alr: \bm w \mapsto \bm \wtilde$:
\[
    \wtilde_j = \log \frac{w_j}{w_H}, \quad j=1, \dots, H-1
\]
and its inverse, $\bm w =\alr^{-1}(\bm \wtilde ):=   
\mathcal{C}(\e^{\wtilde_1}, \ldots, \e^{\wtilde_{H-1}}, 1)$, 
that is 
\begin{equation}
w_j = \frac{\e^{\wtilde_j} }{1+\sum_{h=1}^{H-1}\e^{\wtilde_h}}, \ j=1,\ldots, H-1, \quad w_H= 1-\sum_{h=1}^{H-1} w_h= \frac{1}{1+\sum_{h=1}^{H-1}\e^{\wtilde_h}} \ .
\label{eq:inv_alr}
\end{equation}
Observe that both maps are linear, i.e., for any   $\bm w_1, \bm w_2 \in S^H$, $\tilde{\bm w_1}, \tilde{\bm w_2} \in \mathbb{R}^{H-1}$, $\alpha \in \mathbb{R}$,
\begin{align*}
    \alr(\bm w_1 \oplus \bm w_2) = \alr(\bm w_1) + \alr(\bm w_2), &\quad
    \alr(\alpha \odot \bm w_1) = \alpha \ \alr(\bm w_1) \\
    \alr^{-1}(\tilde{\bm w_1} + \tilde{\bm w_2}) = \alr^{-1}(\tilde{\bm w_1}) + \alr^{-1}(\tilde{\bm w_2}), &\quad
    \alr^{-1}(\alpha \tilde{\bm w_1}) = \alpha \odot \alr^{-1}(\tilde{\bm w_1}). 
\end{align*}
The alr transformation is often applied in the context of compositional data analysis, where statistical inference for data in the simplex has been pioneered by \cite{aitchison1986compositional}. In particular, this
 map was used in \cite{atchison1980logistic} to define a new distribution on the simplex, the logistic-normal distribution.
Formally, we say that $\bm w=(w_1,\ldots,w_{H-1}$, $w_H:=1-\sum_{h=1}^{H-1}w_h)^T \in S^H$ follows the logistic-normal distribution of parameters $\bm\mu, \Sigma$ for $\bm \mu \in \mathbb{R}^{H-1}$, and $\Sigma$ a positive definite $(H- 1) \times (H - 1)$ matrix if 
$$
\bm \wtilde = \alr(\bm w) =\left( \log\frac{w_1}{w_H},\ldots, \log\frac{w_{H-1}}{w_H}\right)^T \sim \calN_{H-1}(\bm\mu, \Sigma)
$$
where $ \calN_{H-1}(\bm\mu, \Sigma)$  denotes the $(H-1)$-dimensional Gaussian distribution with mean $\bm\mu$ and covariance matrix $\Sigma$.
The logistic-normal distribution offers a rich way to model data embedded on the simplex and is particularly suited for our application.
Although moments of this distribution exist, their expression is not available analytically. However, when modeling data in the simplex, one is usually more interested in the pairwise ratios of the components than on the values of the components themselves.
In turn, these expected values and covariances are available analytically and given by
\begin{align*}
    \E \left[\log \frac{w_i}{w_j} \right] = \mu_i - \mu_j,  %\\ 
  \quad   \Cov \left(\log \frac{w_i}{w_j}, \log \frac{w_l}{w_k} \right) = \Sigma_{il} + \Sigma_{jk} - \Sigma_{ik} - \Sigma_{jl} 
\end{align*}
where $\Sigma_{il}$ denotes the $(i,l)$-element of the matrix $\Sigma$.

\section{The logistic MCAR distribution}
\label{sec:logisticMCAR}

In this section, we introduce and describe a joint distribution for a collection of
vectors in the simplex $\bm w_1, \ldots, \bm w_I \in S^H$, reflecting the areal proximity structure in the distribution.
For each pair of areas $i$ and $j$, $g_{ij} \in [0, 1]$ indicates the amount of spatial   proximity   between them.
In the rest of the paper we assume $g_{ij} = 1$ if $i$ and $j$ are neighbors, i.e. the areas share at least a border, and $g_{ij} = 0$
otherwise, but we  could consider more general settings. By definition, $g_{ii}=0$ for all $i$.  The matrix $G=[g_{ij}]_{i,j=1}^I$ is called the proximity matrix and we assume it known.
It will be useful, for our analyses, to identify the matrix $G$ with a graph,
 whose nodes are  denoted by indexes $1, \ldots, I$ and the links are given by the $g_{ij}$'s, i.e. there is a link between nodes $i$ and $j$ if, and only if, $g_{ij}=1$.
We define the joint distribution of $\bm w_1, \ldots, \bm w_I$ introducing the transformed vectors $\bm\wtilde_i:=\alr(\bm w_i)$, $i=1,\ldots,I$ and assuming a joint Gaussian conditional autoregressive distribution for $(\bm\wtilde_1,\ldots, \bm\wtilde_I)$.

Conditionally autoregressive (CAR) models are a special case of Markov random fields.
In general, if $\{X_1, \ldots, X_n\}$, with $X_i\in \R$, is a set of random variables, to define a CAR model over
$X_1, \ldots, X_n$, one usually starts by assigning the conditional distribution of
each $X_i$ given all the others $ X_{-i} :=\{ X_1, \ldots, X_{i-1}, X_{i+1}, \ldots X_n\}$.
The set of conditional distributions, under assumptions, identifies the unique joint distribution of $(X_1, \ldots, X_n)$. The class of CAR models is large; see further detail in  \cite{besag1974spatial}, \cite{cressie1992statistics}, \cite{cressie1993spatial}, \cite{kaiser2000construction}, \cite{cressie2015statistics} and
references therein, just to include a few papers.  

We generalize the univariate CAR model in \cite{leroux2000estimation}  assuming the following multivariate conditionally autoregressive (MCAR) 
model:
\begin{equation}\label{eq:mcar_tilde}
    \bm \wtilde_i \mid \bm \wtilde_{-i}, \Sigma, \rho \sim \calN_{H-1}\left( \frac{\rho \sum_{j=1}^I g_{ij} \bm \wtilde_{j} + (1 - \rho) \bm \mtilde_i}{\rho \sum_{j=1}^I g_{ij} + 1 - \rho}, \frac{\Sigma}{\rho \sum_{j=1}^I g_{ij} + 1 - \rho} \right), \ i=1,\ldots,I,
\end{equation}
where $\Sigma$ is a definite positive $(H-1)\times (H-1)$ matrix and $ \bm \mtilde_i \in \R^{H-1}$ for all $i$. When $H-1=1$, \eqref{eq:mcar_tilde} gives  the prior proposed in \cite{leroux2000estimation}.
If $\rho \in (-1, 1)$,  the joint distribution is well defined and unique \citep[for a proof, see][]{gelfand2003proper}. From \eqref{eq:mcar_tilde} we have that $\bm \wtilde = vec(\bm \wtilde_1 , \ldots, \bm \wtilde_I)$, the vectorization of the weights, is such that 
\begin{equation}
     \bm \wtilde \sim \calN_{I (H-1)} \left( \bm \mtilde, \left((F - \rho G) \otimes \Sigma^{-1} \right)^{-1} \right)
    \label{eq:joint_car}
\end{equation}
where $\otimes$ denotes the Kronecker product, $\bm \mtilde = vec(\bm \mtilde_1 , \ldots, \bm \mtilde_I
)$ and $F = \text{diag}( \allowbreak \rho \sum_j g_{1j} + 1-\rho, \ldots, \rho \sum_{j} g_{Ij} + 1-\rho)$. The matrix 
$A ^{-1}(G,\rho):=(F - \rho G)= \rho \left( diag(G\bm 1_I)-G\right) +(1-\rho)\mathbb{I}_I $  in \eqref{eq:joint_car}, where $\bm 1_I \in \R^I$ denotes the vector of ones and $\mathbb{I}_I $ denotes the $I\times I$ identity matrix, has a key role here. When $\rho = 1$, \eqref{eq:mcar_tilde}  reduces to the intrinsic CAR model, and the joint density of $(\bm\wtilde_1,\ldots, \bm\wtilde_I)$ is improper. If $\rho=0$, 
the $\bm \wtilde_i$'s are independent.  See below for further properties of $A (G,\rho)$. 

We say that the sequence of vectors $\bm w_1, \ldots, \bm w_I$ follows a logistic multivariate CAR distribution of parameters $\rho$ and $\Sigma$ on a graph $G$ if the transformed variables $(\bm \wtilde_1, \ldots, \bm \wtilde_I)$, $\bm \wtilde_i = \alr(\bm w_i)$, follow the MCAR model in \eqref{eq:mcar_tilde} (or \eqref{eq:joint_car}). 
We write  $(\bm w_1, \ldots, \bm w_I) \sim \text{logisticMCAR}(\bm \mtilde$, $\rho, \Sigma; G)$.

One key aspect is the relation that \eqref{eq:mcar_tilde} induces over the vectors 
on the simplex rather than on their alr-transformation.
This is made clear by the following proposition. 

\begin{prop}
\label{prop:cond_means_s}
If $(\bm w_1, \ldots, \bm w_I) \sim \text{logisticMCAR}( \bm \mtilde, \rho, \Sigma; G)$, then, for any $i=1,\ldots,I$,
%    Let the conditional distribution of $ \bm \wtilde_i$ be as in Equation \eqref{eq:mcar_tilde}, and assume $g_{ij} = 1$ if $i$ and $j$ are neighbors and
%    $g_{ij} = 0$ otherwise, call $\bm m_i = alr^{-1}(\bm \mtilde_i)$, 
%    then  
\begin{equation}\label{eq:cond_mean_s}
    \E \left[\log \frac{w_{il}}{w_{ik}} \mid \bm w_{-i} \right] = \log
    \left(\left(\frac{m_{il}}{m_{ik}}\right)^{1-\rho} \prod_{j \in U_i} \left(\frac{w_{jl}}{w_{jk}}\right)^\rho \right)^{(\rho |U_i| + 1 - \rho)^{-1}} \quad l,k=1,\ldots,H
    \end{equation}
    where $U_i = \{ j: g_{ij}>0\}$, $\left| U_i \right| = \sum_j g_{ij}$ and $\bm m_i=(m_{i1},\ldots,m_{iH})$, with $\bm m_i = alr^{-1}(\bm \mtilde_i)$. 
    
\smallskip\noindent
Proof: see Appendix~\ref{sec:app_proofs}.
\end{prop}
There are several immediate but interesting properties of \eqref{eq:cond_mean_s}.
First of all, if $\rho = 1$, \eqref{eq:cond_mean_s} means that the expected value of (the logarithm of) the ratios between the components of $\bm w_i$ is equal to (the logarithm of) the geometric mean of the corresponding ratios of the components of the vectors $\bm w_j$ nearby.
If $\rho = 0$,  the right hand side of \eqref{eq:cond_mean_s} reduces to $\log (m_{il} / m_{ik})$, which is to be expected since, in this case, the $\bm w_i$'s would not be spatially correlated. Instead, in case $0 < \rho < 1$, which we assume throughout the paper (see Section~\ref{sec:themodel}), we can interpret  the right hand side of \eqref{eq:cond_mean_s}
as a weighted mean on the simplex, according to Aitchison geometry, of two components:
the first component $m_{il} / m_{ik}$ corresponding to the mean $\bm m$ 
and the second $\prod_{j \in U_i} \left(w_{jl} / w_{jk}\right)$ taking into account the spatial dependence.
In other words, Proposition \ref{prop:cond_means_s} provides  
the same interpretation of \eqref{eq:mcar_tilde}  but for ratios between components of the anti-transformed 
vectors in the simplex, if we look at them through the Aitchison geometry.

Starting from the joint distribution in \eqref{eq:joint_car}, we can  also study the
marginal covariance of  $\bm w_i, \bm w_j$ in $(\bm w_1, \ldots, \bm w_I) \sim \text{logisticMCAR}( \bm \mtilde, \rho, \Sigma; G)$ for $i\neq j$. 
We point out that the matrix $A (G,\rho)^{-1}$, introduced above, is 
%A key role is played by $A = (F - \rho G)^{-1}$; being $(F - \rho G)$ 
a strictly diagonal dominant matrix (i.e. for each row, the absolute value of the diagonal entry is larger than or equal to the sum of the absolute values of the off-diagonal entries in that row) with negative off-diagonal entries, 
%it is a monotone matrix, i.e. $Av \geq 0$ implies $v \geq 0$ (element-wise) 
and, hence, its inverse $A(G,\rho)$ has elements which are all positive.
\begin{prop}
\label{prop:marg_cov}
If $(\bm w_1, \ldots, \bm w_I) \sim \text{logisticMCAR}( \bm \mtilde, \rho, \Sigma; G)$, then 
%If $\wtilde_1, \ldots, \wtilde_I$ is distributed as \eqref{eq:joint_car}, then:
 \begin{align*}
        \Cov \left(\log \frac{w_{il}}{w_{im}}, \log \frac{w_{jl}}{w_{jm}} \right) &= A_{ij}\left(\Sigma_{ll} - 2\Sigma_{lm} + \Sigma_{mm}\right) \quad i,j = 1, \dots, I \quad 
        l,m =1, \dots, H -1 \\
    	\Cov \left(\log \frac{w_{il}}{w_{iH}}, \log \frac{w_{jl}}{w_{jH}} \right) &= A_{ij}\Sigma_{ll}  \quad i,j = 1, \dots, I \quad 
    	l =1, \dots, H -1  
 \end{align*}  
    In particular, $\Cov \left(\log \frac{w_{il}}{w_{im}}, \log \frac{w_{jl}}{w_{im}} \right) = 0$ if areas $i$ and $j$ belong to different connected graph components of the graph $G$.
    
    \smallskip\noindent
Proof: see Appendix~\ref{sec:app_proofs}.
\end{prop}

Observe that the logisticMCAR distribution of  $(\bm w_1, \ldots, \bm w_I)$ is
not exchangeable, i.e. it is not true that $\Law(\bm w_1, \ldots, \bm w_I)$ and $\Law(\bm w_{\pi(\{1\})}, \ldots, \bm w_{\pi(\{I\})})$  are equal for any 
$(\pi(\{1\}), \ldots, \pi(\{I\}))$  permutation of $(1,\ldots,I)$.
Here, as in the rest of the paper, the
distribution of a random element $y$ is denoted by $\Law(y)$.
Nonetheless, the logisticMCAR distribution 
induces exchangeable priors on all the fully connected
components of the graph $G$.

The logisticMCAR distribution shares the same limitation as
the logistic-normal one, i.e. moments are not available in closed-form expressions.  
In Appendix~\ref{sec:app_MC} we report an extensive Monte Carlo (MC) simulation where we compute the covariance between different components of the vectors of weights
and we draw a comparison between the logisticMCAR and the Dirichlet distributions.

\section{Spatially dependent mixture models}
 
We return to the problem of formalizing a  Bayesian model for $I$ groups of data $(\bm y_1, \ldots, \bm y_I)$, $\bm y_i = (y_{i1}, \ldots, y_{iN_i})$, $i=1,\ldots,I$. As mentioned at the beginning of Section~\ref{sec:logisticMCAR}, we assume that each vector $\bm y_i$ is associated to an area $i$ and that
for each pair of areas $i$ and $j$, $g_{ij} = 1$ if $i$ and $j$ are neighbors and $g_{ij} = 0$ otherwise.

\subsection{The finite mixture model with spatially dependent weights}
\label{sec:themodel}
Let the proximity matrix $G=[g_{ij}]_{i,j=1}^I$ be fixed. 
We assume that $\bm y_1, \ldots, \bm y_I$, conditioning to $\bm w_1,\ldots,\bm w_I$ and $\bm \tau$, are independent and that, for each $i=1,\ldots,I$,
\begin{align}
    y_{ij} \mid \bm w_i, \bm \tau & \iid \sum_{h=1}^H w_{ih} \calN(\cdot \mid \tau_h) \quad j=1,\ldots,N_i,  \label{eq:lik} \\
    \tau_h & \iid P_0 \quad h=1,\ldots,H \label{eq:prior_tau} \\
    (\bm w_1, \ldots, \bm w_I)\mid \rho, \Sigma & \sim \text{logisticMCAR}(\bm \mtilde, \rho, \Sigma; G) \label{eq:prior_weights} \\
    \Sigma & \sim \text{Inv-Wishart} (\nu, V) \label{eq:prior_Sigma} \\
    \rho & \sim \pi(\rho) \label{eq:prior_rho}
\end{align}
where $\bm w_i=(w_{i1},\ldots,w_{iH})^T\in S^{H-1}$ (see \eqref{eq:def_simplex}) and $\bm \mtilde = vec(\bm \mtilde_1 , \ldots, \bm \mtilde_I
)\in \R^{I(H-1)}$.
As often considered, we study the case where the kernel in the mixture \eqref{eq:lik} is the Gaussian density with  mean $\mu_h$ and variance $\sigma_h^2$, so that  
$\tau_h = (\mu_h, \sigma_h^2)$ and $P_0$ is a probability distribution over $\Theta=\mathbb{R} \times \mathbb{R}^+$.  Specific choices of $P_0$ are discussed in Sections \ref{sec:simulation} and \ref{sec:airbnb}.
We consider independent marginal priors for $\rho$ and $\Sigma$. Moreover, the support of the prior of $\rho$ is typically assumed to be $(0,1)$ to induce the similarity of spatial neighbors  \citep[see, for instance,][Section 4]{gelfand2003proper}.

Model \eqref{eq:lik} - \eqref{eq:prior_rho} assumes that each group of data $\bm y_i$ is modeled as a (finite) mixture of  Gaussian kernels.
Specifically, observations within each group are i.i.d given the weights 
and the atoms of the mixtures, while 
conditionally to all the mixture weights $\bm w_1, \ldots, \bm w_I$, observations in different groups are independent.
All the $I$ mixtures share the same set of atoms $\tau_1, \ldots, \tau_H$ 
, which are assumed i.i.d from the base measure $P_0$,  a continuous distribution on 
$\Theta = \mathbb{R} \times \mathbb{R}^+$.
The dependence between mixtures in different areal units is induced only by the prior on the mixture weights.

%\MPtext{As mentioned in the Introduction,  we define a prior for 
%$(\bm w_1, \ldots, \bm w_I)$, allowing weights associated to close areas to be more similar than weights associated to areas farther away, through the logistic transformation of a Gaussian CAR model. The idea is not new in the literature, and the prior for the mixture weights of area-dependent densities in \cite{jo2017dependent} is 
%closely related to our prior. 
%We discuss the differences between the two priors in Section~\ref{sec:comparison} and compare their features by fitting simulated data to the two models in Section~\ref{sec:nongaussian}.}
%\MPnote{il pezzo qua in blu spezza abbastanza. penso si possa togliere oppure spostarlo a fine sottosezione}

In order to derive a Gibbs sampler for our model, we introduce the latent variables $s_{ij}$, one for each observation, indicating to which component of the mixture  observations are allocated to, and rewrite \eqref{eq:lik} as
\begin{align}
    y_{ij} \mid s_{ij} = h, \tau_h & \ind  \calN(\cdot \mid \tau_h)
    \quad j=1,\ldots,N_i,\  i=1,\ldots,I \label{eq:lik_latent}\\
    p(s_{ij} = h \mid \bm w_i) & = w_{ih} \quad h=1,\ldots,H
     \label{eq:cluster_allocs}
\end{align}
%Note that the latent variable $s_{ij}$ in \eqref{eq:cluster_allocs} denotes the component (in $\{1,\ldots,H\}$) $y_{ij}$ in \eqref{eq:lik_latent} has been allocated to. 
A component in the mixture is said \textit{empty} if it has not been allocated to any observation. Here, and in the whole paper,  \textit{cluster} denotes any allocated component and the number of clusters is the number of allocated components. It is clear from \eqref{eq:lik_latent}-\eqref{eq:cluster_allocs} that the allocated and empty components, as well as the number of clusters, are random variables, with marginal prior distributions
 induced by our model. 

We complete the specification of our model by adopting a marginal prior on 
$\bm \mtilde$ that encourages sparsity in the mixtures.
We discuss this choice in detail in the next Section~\ref{sec:prior_mtilde}.

\subsection{Sparse mixtures via a prior on $\bm \mtilde$}\label{sec:prior_mtilde}

Generally,  a sparse mixture is obtained when the number of clusters is smaller than the total number of components $H$.
There are two well-known strategies to obtain sparse mixtures in the Bayesian context.
The first one assigns a prior on the weights that forces them to be 
stochastically decreasing, so that the \virgolette{last} weights are
very small and the corresponding mixture components are seldom allocated.
The alternative strategy consists in assigning a prior for the weights that concentrates
its mass around the edges of the simplex in a symmetric way, as it is the case
of the sparse Dirichlet distribution,  i.e. a Dirichlet distribution
with all the parameters equal to $\alpha$, with $0<\alpha<1$.
In the latter case, there is no preferential ordering of the weights and any 
mixture component could be allocated.
We think that the first approach might not fit spatial applications, in particular 
when the proximity graph $G$ has disconnected components, since assuming   decreasing weights for all the mixtures  would force data from two disconnected components to be always sampled from the few components with larger weights, and hence to behave always similarly.  

Here, we show how we can mimic the sparse Dirichlet distribution for the weights, by assuming a suitable prior on
parameters $\bm \mtilde_i$'s in our model.
We start by observing that in the mixture model \eqref{eq:lik} for the   $i$-th area,  if coordinate values in the vector $\bm \mtilde_i$ 
in \eqref{eq:prior_weights}  are very different among each other, this would force some components $h$ in \eqref{eq:lik} to be more often allocated than others, being their weights larger than the others (in mean).
Hence, we induce  \virgolette{symmetric} sparsity in our marginal prior for the weights by assuming 
$\bm\mtilde_i \sim \calN_{H-1}(\bm 0, \eta^2 \mathbb{I})$.
Observe that, since the distribution of $\bm \mtilde_i$ is centered in $\bm 0$
and isotropic, we are not forcing, marginally, any specific ordering on the weights.

To understand  the role of $\eta^2$, let us consider an illustrative example when $H=3$ and $I=1$. Let $\bm m_1 = \alr^{-1} \bm \mtilde_1$ and consider  $d_{12} = (\log(m_{11} / m_{1H}) - \log(m_{12} / m_{1H}))^2$, which corresponds to the distance between $m_{11}$ and $m_{12}$ in the Aitchison geometry. We may consider $d_{12}$ as a plug-in estimator of the distance between $w_{11}$ and $w_{12}$.
The largest values of $d_{12}$ are obtained when one between $m_{11}$, $m_{12}$ and $m_{1H}$ is  approximately 1 and the others are close to zero. 
Moreover, from $\bm\mtilde_1 \sim \calN_{2}(\bm 0, \eta^2 \mathbb{I})$, we have that $d_{12} /(2 \eta^2)$ has chi-squared distribution  with one degree of freedom.
Hence the random variable $d_{12}$ is stochastically increasing with $\eta^2$.

This feature holds also for larger values of $H$ as shown in Figure~\ref{fig:num_components_comparison},  where the behavior of $\bm w_i$, for different values of $\eta^2$,  is illustrated. We conclude that $\eta^2$ is a sparsity tuning parameter and sparsity of the $\bm w_i$'s is obtained for larger values of $\eta^2$.
Note that this is the opposite behavior of other sparsity priors, such as the double exponential or the horseshoe \citep{bhadra2019lasso}, where a distribution with significant mass near zero is assumed.
Because our parameters are transformed through the logistic map \eqref{eq:inv_alr}, assuming a prior concentrated in zero for $\bm \mtilde$ would result in a prior concentrated on $(1/H, \ldots, 1/H)$ for $\bm w$,  of course this being far from sparsity. 

%\mariosub{
%In addition, since for large values of $\eta^2$ a small number of components 
%in $\mtilde_i$ will be significantly larger than all the others,
%under the marginal logisticMCAR prior for the weights $\bm w_i$ in a single 
%area $i$,  the number of clusters will have a prior distribution that is highly peaked 
%on values near 1; see also Figure~\ref{fig:num_components_comparison}(b).
%On the other hand, when $\eta^2$ is  very small, all the components in $\bm \mtilde_i$
%are close to $0$, implying that all the weights in \eqref{eq:lik} are close to $1/H$ (see \eqref{eq:inv_alr}) and consequently a large  prior number of clusters in each finite mixture.}{}

Assuming each $\bm \mtilde_i$ in the prior specification \eqref{eq:mcar_tilde} to be random and i.i.d would imply that, a priori,  
the mixture weights could be extremely different in neighboring areas. 
However, this is not what we aim at modeling in applications, where we typically
assume that data in close areal units  follow similar distributions.  
On the other hand, assuming $\bm \mtilde_1=\ldots=\bm \mtilde_I$ seems overly restrictive,
since we would loose the property that two connected graph components in 
$(\bm \wtilde_1,\ldots,\bm \wtilde_I)$
are independent under CAR distributions when marginalizing out the only shared parameter $\bm \mtilde_1$. 
%\bale Se usiamo questa generalizzazione negli esempi, vale la pena di numerare le formule e poi richiamarle. \eale
Hence %, as a trade-off between these opposite demands, 
we propose to extend the logisticMCAR$(\bm \mtilde, \rho, \Sigma; G)$ in  \eqref{eq:prior_weights} assuming %in $\bm \mtilde=(\bm \mtilde_1,\ldots,\bm \mtilde_I)$:
\begin{equation}
    \label{eq:m_connected}
    \{\bm\mtilde_i = \bm\mtilde_j = \bm\mtilde_{C_m} \text{ iff } i, j \in C_m \text{ for some } m\}
\end{equation} 
where $C_1, \ldots, C_k$ denote the connected components of graph $G$, i.e. 
 all the parameters $\bm\mtilde_i$s are assumed  common within each connected component. For such parameters, we assume
\begin{equation}
    \label{eq:m_connected_prior}
    \bm \mtilde_{C_1}, \ldots \bm\mtilde_{C_k} \iid \calN_{H-1}(\bm 0, \eta^2 \mathbb{I}).
\end{equation}

\subsection{Comparison with competitor models}
\label{sec:comparison}

As mentioned in the Introduction,  we have defined a prior for 
$(\bm w_1, \ldots, \bm w_I)$, allowing weights associated to close areas to be more similar than weights associated to areas farther away, through the logistic transformation of a Gaussian CAR model. The idea is not new in the literature, and the prior for the mixture weights of area-dependent densities in \cite{jo2017dependent} is 
closely related to our prior. 
We discuss the differences between the two priors in  this section and we further compare their features by fitting simulated data to the two models in Section~\ref{sec:nongaussian}.

We briefly introduce the class of spatially dependent species sampling mixtures in \cite{jo2017dependent}, who define the weights in the mixtures to be spatially dependent,  modeling them from a Gaussian CAR distribution, as we do. 
Their focus is on geo-referenced data (with multiple observations in each geographic location), and they propose two different CAR specifications, namely the Mercer CAR and the Clayton-Kaldor CAR \citep{clayton1987empirical} priors. 
Since it is not straightforward  to extend the Mercer CAR formulation to areal data as  it requires the computation of a geographical distance rather than defining a proximity matrix, we only consider the Clayton-Kaldor CAR species sampling model in \cite{jo2017dependent} for comparison. 
We have shown in Section~\ref{sec:prior_mtilde} that our marginal prior 
can mimic the sparse Dirichlet distribution by assuming $\bm \mtilde$   
in the logisticMCAR$(\bm \mtilde, \rho, \Sigma; G)$ to be random.
Below, we discuss how sparsity is obtained also in the spatially dependent species sampling model in \cite{jo2017dependent}, but only in some sort of \virgolette{asymmetric} manner, and how this impacts the modeling of different connected components in the graph $G$.

In the following, we refer to the prior in \cite{jo2017dependent} as \emph{CK-SSM}. Instead of jointly modeling the transformed weights in each location, \cite{jo2017dependent} assume independent univariate CAR model for (a transformation of) the weights associated to each component of the mixture in the different areas. Recall that they assume a mixture with infinite components, i.e., $h=1,2,\ldots$. With our notation, let $\bm \nu_h = (w_{1h}, \ldots, w_{Ih})$, then the CK-SSM prior for $\bm \nu_h$ is
\begin{equation}\label{eq:priorweights_Jo}
    \widetilde{\bm \nu}_h  \ind \mathcal{N}_I(\widetilde{\theta}_h, \tau^2 (I - \rho G)) \ \ h=1,2 \ldots, \quad \nu_{ih} = w_{ih} = \frac{\e^{\widetilde{\nu}_{hi}}}{\sum_j \e^{\widetilde{\nu}_{ji}}} \ \ i=1,\ldots,I
\end{equation}
In order to guarantee  that the denominator of the fraction in  \eqref{eq:priorweights_Jo} 
 is finite, \cite{jo2017dependent} assume  that $\widetilde{\theta}_h$ is a vector with all components equal to $\log\{1 -(1+ \e^{b-ah})^{-1}\}$, $a$ and $b$ being positive hyperparameters.  In force of that, the weights $\widetilde{\nu}_{hi}$ are stochastically decreasing with $h$ for each area $i$. This ordering is preserved by the exponential and normalization transformations,  so that $w_{ih}$ will be stochastically decreasing with $h$ as well, for each $i$. Note also that \eqref{eq:priorweights_Jo} makes $\bm w$ non-identifiable.

\begin{figure}[t]
	\centering
	\includegraphics[width=0.7\linewidth]{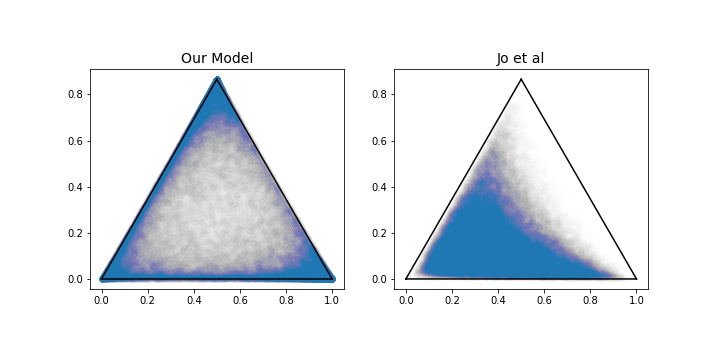}
	\vspace{-5mm}
	\caption{Scatterplots of $N=100,000$ MC draws (in blue) from the marginal priors of the weights in one single area with  $H=3$, under our logisticMCAR prior (left) and 
\eqref{eq:priorweights_Jo} in \cite{jo2017dependent} (right). White/gray areas
represents low-density zones and dark blue zones high-density ones.}
	\label{fig:our_vs_jo_model}
\end{figure}

We start by considering $I=1$ area and drop the subscript $i$. As in \cite{jo2017dependent}, we truncate \eqref{eq:priorweights_Jo} to the first $H$ terms  for computation.
Figure~\ref{fig:our_vs_jo_model} shows a comparison of the marginal priors of the weights in the mixture \eqref{eq:lik} with $H=3$, under our logisticMCAR prior and 
\eqref{eq:priorweights_Jo} introduced in \cite{jo2017dependent}. 
In particular, for our logisticMCAR$(\bm \mtilde, \rho, \Sigma; G)$ prior we have assumed
    $\bm \mtilde \sim \calN_2(\bm 0, 9 \mathbb{I})$,
    $\bm \wtilde \sim \calN_2(\bm \mtilde, \mathbb{I})$, 
that is \eqref{eq:joint_car} with $\Sigma=\mathbb{I}$, 
while we fix $\widetilde{\theta}_h= \log\{1 -(1+ \e^{1-h})^{-1}\}$, $h=1,\ldots,H$, in \eqref{eq:priorweights_Jo} as in  \cite{jo2017dependent} ($a=b=1$) and $\tau^2=1$. 

We compare the priors via $N=100,000$ MC draws. 
Figure~\ref{fig:our_vs_jo_model} shows the scatterplots of the draws of the two marginal priors.
In particular, the draws from our prior (left panel) recover  the \virgolette{sparse}  
symmetric Dirichlet prior with all  parameters equal to $\alpha < 1$; the draws are symmetrically concentrated around
the edges of the simplex,  and give
significant mass to locations near the vertexes.
On the other hand, the draws from the CK-SSM prior clearly show asymmetry in favor of the first component, also giving negligible mass  to neighborhoods of the vertices.
When the number $H$ of components in the mixture \eqref{eq:lik} is larger, we can compare the priors via two functionals by computing 
($i$) the number of 
 active components ($H^{(a)}$), that we define as the components associated to  weights greater than
$0.01$, i.e. the cardinality of the set $\{h: \ w_h > 0.01\}$,
and ($ii$) the probability for each component of the vector $\bm w\in S^H$ 
to be greater than the threshold  $0.05$.  
\begin{figure}[h]
	\centering
	\begin{subfigure}[c]{\textwidth}
    	\centering
    	\includegraphics[height = 4 cm]{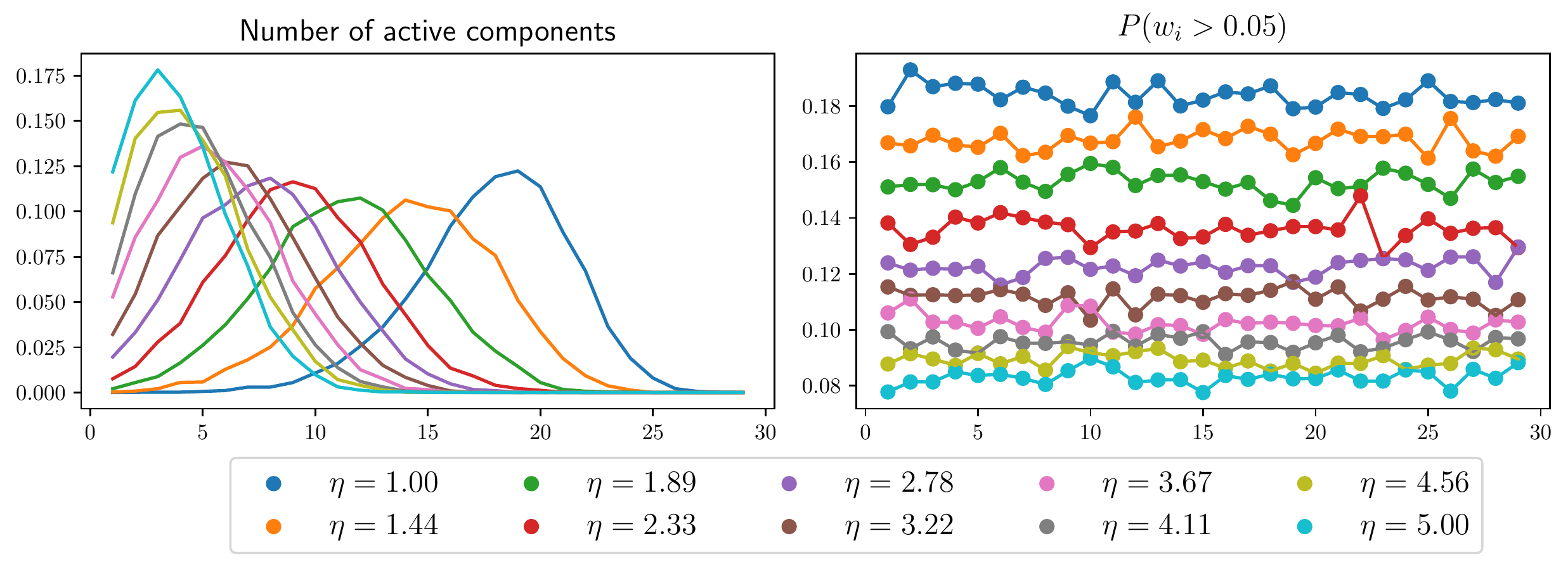}
    	\caption{Our prior for different values of $\eta^2$}
    	\label{fig:num_components_ours}
    \end{subfigure}
   
	\begin{subfigure}[c]{\textwidth}
		\centering
		\includegraphics[height = 4 cm]{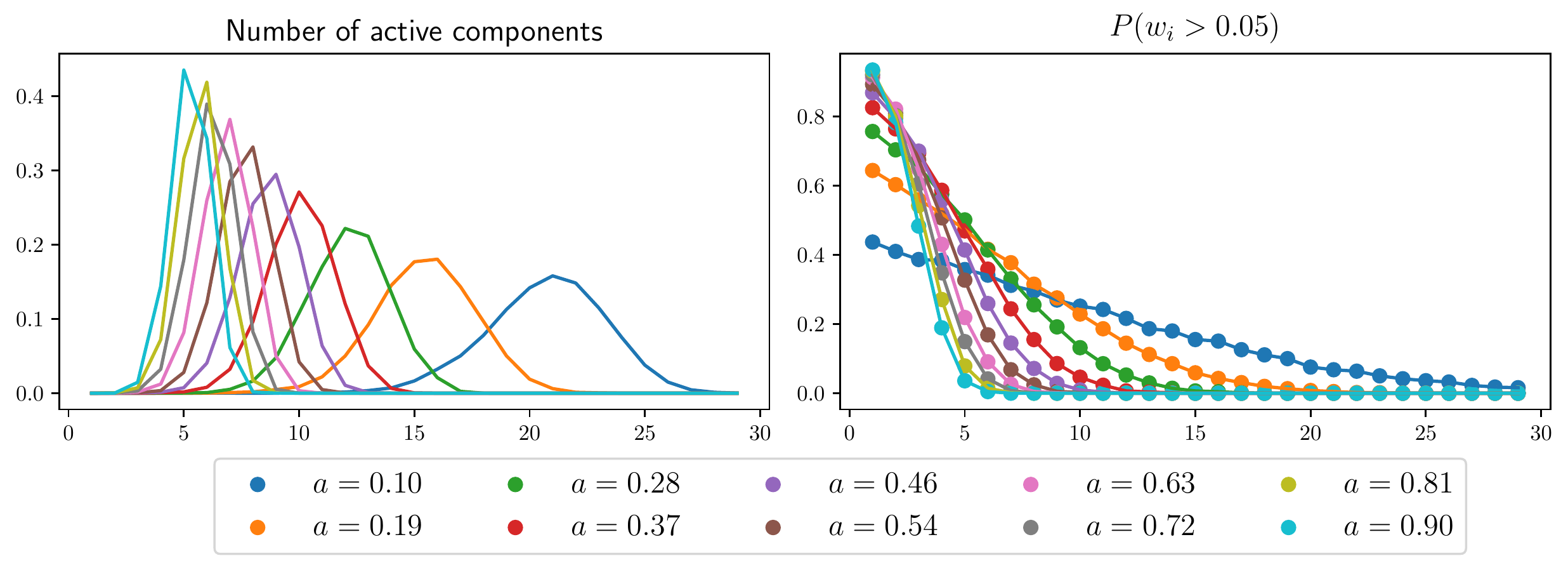}
		\caption{The prior in \cite{jo2017dependent} for different values of $a$ and $b=0.5$}
		\label{fig:num_components_jo}
	\end{subfigure}
\caption{Prior distribution of the number of active components (left) and
    the probability for $w_h$ to be greater than $0.05$ (right) under our prior (top row) and prior \eqref{eq:priorweights_Jo} in \cite{jo2017dependent} (bottom row). Here $H=30$.}
\label{fig:num_components_comparison}
\end{figure}
We fix $H=30$ and, simulating $N=10,000$ MC draws as before, we plot the marginal  prior distributions of these functionals under
our logisticMCAR prior (Figure~(\ref{fig:num_components_comparison}a)) and 
\eqref{eq:priorweights_Jo} in \cite{jo2017dependent} (Figure~(\ref{fig:num_components_comparison}b)), for 
different values of the hyperparameters in the priors. From both left panels, displaying the marginal priors of $H^{(a)}$ (as continuous lines to help seeing the differences), it is clear that the two models may induce different types of prior behaviors. However, when considering the right panels, displaying, for each index $h=1, \ldots, H$, the probability that $w_h > 0.05$,
it is clear that, while under our prior, for each 
degree of sparsity $\eta^2$, 
the probability of inclusion of a single component does not show a preferential ordering, this probability decreases with  $h$ under the  CK-SSM prior.
Going back to the prior in \eqref{eq:m_connected_prior},
observe how this model specification 
gives a major difference with the mixture model in \cite{jo2017dependent}.

 This is particularly relevant if we aim at considering the context where areal units are connected through the graph $G$, but there are at least two different connected 
components,  as we will have in the application in 
Section~\ref{sec:airbnb}.
Intuitively from the discussion above, CK-SSM would still force the different
connected components in the graph to behave similarly, because of the parameter $\bm{\widetilde{\theta}}$ shared by all the mixtures; see \eqref{eq:priorweights_Jo}.
We tested this scenario more in detail by considering a spatial domain subdivided in four areas with two connected components $\{1, 2 \}, \{3, 4\}$. Figure~\ref{fig:prior_dist} shows the total variation distance for $(\bm w_1, \bm w_2)$ and  $(\bm w_1, \bm w_4)$ under the logisticMCAR and CK-SSM priors, having fixed hyperparameters as above and $\rho = 0.95$. It is clear that, as sparsity increases, the distance between $\bm w_1$ and $\bm w_4$ increases under the logisticMCAR but decreases under the CK-SSM,  showing how imposing a sparse behavior in the CK-SSM prior forces similar distributions in disconnected components of the graph. See also Figure~\ref{fig:prior_dens} for a visual representation of draws from the prior distributions.
This effect becomes increasingly more evident as the sparsity in each mixture is increased, as shown in Figure~\ref{fig:num_components_jo}.
On the other hand, our model allows for the required level of sparsity in each
mixture without forcing the different connected components in the graph to behave similarly. For this reason, we believe we have introduced a more flexible model 
for jointly estimate spatially dependent densities than \cite{jo2017dependent}, at least for applications
where different connected components in the graph should exhibit different behaviors.

We will provide comparison also with the  Hierarchical Dirichlet Process (HDP) mixture model in \cite{teh2006hdp}  in Section~\ref{sec:simulation}. To keep the paper self-contained as much as possible, we report the HDP mixture model as follows
%\begin{align}
%    y_{ij} \mid F_i & \iid \int_{\Theta} k(y_{ij} \mid \tau) F_i(d\tau) \label{eq:lik_hdp} \\
%    F_1, \ldots, F_I \mid G &\iid \mathcal{D}_{\alpha G} \label{eq:hdp_prior_first}  \\
%    G & \sim \mathcal{D}_{\beta P_{0}}  
%    \label{eq:hdp_prior_second}
%\end{align}
\begin{equation}\label{eq:hdp}
y_{ij} \mid F_i  \iid \int_{\Theta} k(y_{ij} \mid \tau) F_i(d\tau), \quad
\{F_i\}_{i=1}^I \mid G \iid \mathcal{D}_{\alpha G} \quad G \sim \mathcal{D}_{\beta P_{0}}  
\end{equation}
where $\mathcal{D}_{\beta P_{0}}$ denotes the Dirichlet
measure, i.e. the distribution of a random probability measure that is the Dirichlet process
with measure parameter $\beta P_0$. We assume the kernel $k(\cdot \mid \tau)$ as the Gaussian density on $\mathbb{Y}$ for $\tau=(\mu,\sigma^2)$ as in \eqref{eq:lik}. 
Thanks to the stick-breaking representation of the Dirichlet process, it is possible to rewrite the likelihood in \eqref{eq:hdp} as
\[
    y_{ij} \mid \{w_{ih}\}_{h=1}^{\infty}, \{\phi^*_{ih}\}_{h=1}^{\infty} \iid \sum_{h=1}^{\infty}w_{ih} k(y_{ij} \mid \phi^*_{ih})
\]
where $\phi^*_{ih} \mid G \iid G$ in \eqref{eq:hdp} and $\{w_{ih}\}$ for each $i$ are a sequence
of non-negative weights summing to 1.
Moreover,  since each $F_i$, conditionally to $G$, 
is an independent draw from the Dirichlet process prior with discrete base measure $G$, this yields that all the atoms are shared across all populations. This means that the set of the unique values in $\{\phi^*_{ih}\}_{h=1}^{\infty}$ is equal to the set of unique values in $\{\phi^*_{jh}\}_{h=1}^{\infty}$ for $j \neq i$ and coincides with the set of atoms in $G$.
Hence, denoting by $\{\tau_h\}_{h=1}^\infty$ the atoms in $G$, the HDP mixture model defines a joint probability distribution for random probability measures with the same support points, as in our model. This is 
the motivation to consider the HDP mixture model as the \virgolette{natural competitor} of ours.   

\section{The Gibbs Sampler}
\label{sec:gibbs}
We illustrate a MCMC algorithm to sample from the posterior distribution of
our model \eqref{eq:lik}-\eqref{eq:prior_rho}  and \eqref{eq:m_connected} - \eqref{eq:m_connected_prior}. The state is described by parameters 
 $\bm \tau = (\tau_1, \ldots, \tau_H)$, $(\bm \wtilde_1, \ldots, \bm \wtilde_I)$, where $\bm \wtilde_i = \alr(\bm w_i)$, $i=1,\ldots,I$, $\{s_{ij}\}_{ij}$ ($j=1,\ldots,N_i$) in \eqref{eq:lik_latent}-\eqref{eq:cluster_allocs} and $\bm \mtilde_{C_1}, \ldots \bm\mtilde_{C_k}$ in \eqref{eq:m_connected_prior}.
%\bale Illustriamo l'MCMC di base? Quindi non quello con $\mtilde$ aleatorie?\eale

We use the following notation: given the sequence of vectors $\bm w_1, \ldots, \bm w_I$, we denote by $\bm w_{-i}$ the same sequence where the $i$--th vector  has been removed.
Given a single vector $\bm w_i$, we denote by $\bm w_{i, -h}$ the same vector where the $h$-component as been removed.
Finally, for the matrix $\Sigma$, let $\Sigma_{ij}$ denote  the $(i,j)$-element; moreover, $\Sigma_{i}$ denotes its $i$--th row (as a vector) so that $\Sigma_{i, -j}$ denotes the $i$--th row where the $j$--th element has been removed and $\Sigma_{-h, -k}$ denotes the $(H-2) \times (H-2)$ matrix there the $h$--th row and $k$--th column have been removed.

There are two \virgolette{non-standard} steps in the Gibbs sampler:
the update of the transformed weights $\bm \wtilde_i$ and the update
of their means  $\bm \mtilde_{C_1}, \ldots \bm\mtilde_{C_k}$.
Here, we only describe the full conditionals of each $\bm \wtilde_i$.
The full conditional of $\bm \mtilde_{C_i}$  is a multivariate Gaussian distribution. See Appendix~\ref{sec:app_Gibbs} for more detail on it,  together with the other standard full conditionals.

We begin by writing the full conditional for $\wtilde_{ih}$, for each $i$ and $h$, as
\begin{equation}
        \Law(\wtilde_{ih} \mid \bm \wtilde_{-i}, \bm \wtilde_{i, -h}, rest) \propto \pi (\wtilde_{ih} \mid \bm \wtilde_{-i}, \bm \wtilde_{i,-h}, \rho, \Sigma) \,\Law(\wtilde_{ih} \mid \bm s_i, \bm \wtilde_{i, -h}) \label{eq:full_cond_weights}
\end{equation}
where $\bm s_i=(s_{i1},\ldots, s_{iH})^T$. 
The conditional prior $\pi (\wtilde_{ih} \mid \bm \wtilde_{-i}, \bm \wtilde_{i,-h}, \rho, \Sigma)$ can be derived from \eqref{eq:joint_car} conditioning with respect to the other components of the vector $\bm \wtilde_i$; we find 
\[
    \pi (\wtilde_{ih} \mid \bm \wtilde_{-i}, \bm \wtilde_{i,-h}, \rho, \Sigma) = \calN(\mu_{ih}^*, \Sigma_{ih}^*),
\]
where ${\displaystyle \mu_{ih}^* = \mu_{ih} + \Sigma_{h, -h} \Sigma^{-1}_{-h,-h}(\bm \wtilde_{i, -h} - \bm \mu_{i, -h})}$ and $  
 \Sigma^*_{ih}=(\rho \sum_{j=1}^I g_{ij} + 1 - \rho)^{-1}$ $ \left(\Sigma_{h,h} \right.$ $\left. - \Sigma_{h,-h}\Sigma^{-1}_{-h,-h}\Sigma_{-h,h}\right) $
%%\begin{align*}
%%    \mu_{ih}^* &= \mu_{ih} + \Sigma_{h, -h} \Sigma^{-1}_{-h,-h}(\b\wtilde_{i, -h} - \bm \mu_{i, -h}) \\
%%    \Sigma^*_{ih} &= \frac{1}{\rho \sum_{j=1}^I g_{ij} + 1 - \rho}\left(\Sigma_{h,h} - \Sigma_{h,-h}\Sigma^{-1}_{-h,-h}\Sigma_{-h,h}\right)
%%\end{align*}
by standard properties of the normal distribution, with $\bm \mu_i = (\rho \sum_{j=1}^I g_{ij} + 1 - \rho)^{-1}(  \rho \sum_{j=1}^I g_{ij} \bm \wtilde_{j} + (1 - \rho) \bm \mtilde_i)$. 
Moreover, using the same data augmentation scheme proposed in \cite{holmes2006}, we write the 
%``likelihood'' 
term 
%(\bale perch\'e si chiama cos\'i? \eale) 
$\Law(\wtilde_{ih} \mid \bm s_i, \bm \wtilde_{i, -h})$ as
\begin{equation*}
    \Law(\wtilde_{ih} \mid \bm s_i, \bm \wtilde_{i,-h}) = \left( \frac{\e^{\eta_{ih}}}{1+\e^{\eta_{ih}}}\right)^{N_{ih}} \left( \frac{1}{1+\e^{\eta_{ih}}}\right)^{N_i - N_{ih}}
\end{equation*}
where $\eta_{ih} = \wtilde_{ih} - C_{ih}$, $C_{ih} =\text{log} \sum_{k \neq h}\e^{\wtilde_{ik}}$ (with $\wtilde_{iH} := 0$) and $N_{ih}$ is the number of observations in area $i$ assigned to component $h$.
%Observe that, again by standard properties of the normal distribution, the set of full conditionals defined in \eqref{eq:cond_prior_comp_x} define the unique joint distribution as in \eqref{eq:joint_car}.

To be able to sample from the full conditional of $\wtilde_{ih}$, we express 
$ \Law(\wtilde_{ih} \mid \bm s_i, \bm \wtilde_{i,-h})$ using an augmentation technique, based on the P{\'o}lya-Gamma distribution. The trick is analogous to  that in \cite{polson2013bayesian}, in this case without covariates.
We describe it in detail in the next paragraphs.

We denote by $\omega \sim PG(b,c)$  a random variable with a P\'olya-Gamma distribution with parameters $b$ and $c$, i.e.
\begin{equation}\label{eq:pg_distrib}
    \omega = \frac{1}{2\pi^2}\sum_{k=1}^{+\infty} \frac{g_k}{(k-1/2)^2+c^2/(4\pi^2)}
\end{equation}
where $g_k\iid Gamma(b,1)$ and $b,c>0$.
The data-augmentation technique based on the P\'olya-Gamma distribution relies on the following integral identity:
\[
    \frac{(\e^\eta)^a}{(1+\e^\eta)^b} = 2^{-b}\e^{(a-b/2)\eta}\int_0^{+\infty} \e^{-\omega \eta^2/2}p(\omega)d\omega
\]
where  $p(\omega)$ is the density of the $PG(b,0)$ random variable.

Taking advantage from the above equality, when introducing the latent variable 
$\omega_{ih} \sim PG(N_i, 0)$, we can derive the following full conditional
for $\wtilde_{ih}$:
\begin{equation}\label{eq:full_cond_wih_pg}
    \Law(\wtilde_{ih} \mid \bm \wtilde_{-i}, \bm \wtilde_{i, -h}, \bm s_i, \rho, \Sigma, \omega_{ih}) = N(\hat\mu_{ih}, \hat \Sigma_{ih})
\end{equation}
where
\begin{align*}
    \hat\mu_{ih} = \left(\frac{\mu_{ih}^*}{\Sigma^*_h} + N_{ih}-N_i/2 + \omega_{ih}C_{ih}\right)\left(\frac{1}{\Sigma^*_{ih}}+\omega_{ih}\right)^{-1}  \quad
    \hat \Sigma_{ih} = \left(\frac{1}{\Sigma^*_{ih}} + \omega_{ih}\right)^{-1} . 
\end{align*}
Moreover, the full conditional of $\omega_{ih}$ can be expressed as
\begin{equation}\label{eq:full_cond_omegaih}
    \Law(\omega_{ih} \mid \bm \wtilde_i) = PG \Big(N_i, \wtilde_{ih}- \log \sum_{k \neq h}e^{\wtilde_{ik}} \Big).
\end{equation}
See Appendix~\ref{sec:app_Gibbs} for the proof of Equations \eqref{eq:full_cond_wih_pg}-\eqref{eq:full_cond_omegaih}.
 
These equations give a two steps Gibbs update for the variable $\wtilde_{ih}$.
Indeed, one can first sample $\omega_{ih}$ from \eqref{eq:full_cond_omegaih} (which depends on $\wtilde_{ih}$)  and secondly update $\wtilde_{ih}$ from \eqref{eq:full_cond_wih_pg} (which depends on $\omega_{ih}$). 
In this way, we are able to make two Gibbs steps in an augmented state space instead of a single Metropolis Hastings step.
There are two reasons why   one should prefer the former algorithm to the latter.
First, the two-Gibbs-steps simulation  avoids the choice of a proposal density for the update, that can be difficult due to the shape of the logistic transformation
applied to the weights.
% As this transformation is extremely nonlinear, one shall
%design a proposal that adapts to the domain region of the transformed weight in order
%to propose reasonable values for the anti transformed weight.
Moreover, using the P\'olya Gamma augmentation trick can be helpful in settings where 
the number of observations in a single
area is not significantly greater than the number of components in the mixture, as we consider in Section~\ref{sec:nongaussian}, scenario II; see
Section~S6.3 of the supplementary material in \cite{polson2013bayesian} for an explanation of this statement. 

%%%%%%%%%%%%%%%%%
% SIMULATION 1
%%%%%%%%%%%%%%%%%

\section{Simulated data} 
\label{sec:simulation}
We consider two simulation studies to illustrate the flexibility of our model; in particular we will see that the model is able to 
exploit spatial dependence between
densities corresponding to close areas.
In the first example, we compare our model (SPMIX) with the 
Clayton-Kaldor Species Sampling Model of \cite{jo2017dependent} (CK-SSM) and the HDP mixture model (see \eqref{eq:hdp}), that we use as a sort of black-box model for density estimation of grouped data. 
%The comparison concerns the estimation of non-gaussian densities. We thus simulate data on a disconnected graph, with each component being associated to a different density. 
% Then we illustrate the estimation errors of the different models in terms of Hellinger distance and Kullback-Leibler divergence.
In the second example we generate data from spatially dependent densities and we check if our model is flexible enough to recover such dependence.

We run the Gibbs sampler for our model \eqref{eq:lik}-\eqref{eq:prior_rho} together with the prior specification 
\eqref{eq:m_connected} - \eqref{eq:m_connected_prior} (see Section~\ref{sec:gibbs} and Appendix~\ref{sec:app_Gibbs}), and the \emph{direct sampler} for the HDP-mixture model in \cite{teh2006hdp}. Both algorithms were coded in \texttt{C++}. 
In addition, we have also implemented the CK-SSM model in \texttt{Stan} \citep{stan} with the prior \eqref{eq:priorweights_Jo}.
All the MCMC chains were run for 10,000 iterations after
discarding the first 10,000 iterations as burn-in, keeping one every five iterations, resulting in a final sample size of 2,000, unless otherwise specified. In all cases, convergence was checked using both visual inspection of the chains and standard diagnostics available in the CODA package.

The base measures for our model, for the HDP-mixture and for the CK-SSM mixture are  assumed all equal (and denoted by $P_0$) to match the models under comparison. Unless otherwise stated, we assume $P_0$ equal to the Normal-inverse-gamma distribution with parameters $\mu_0=0, a=b=2, \lambda=0.1$, i.e.
%\begin{equation*}
$ \mu \mid \sigma^2 \sim \calN \left(\mu_0, \lambda^{-1} \sigma^2 \right)$,  $\sigma^2 \sim IG(a, b)$ 
%\end{equation*}
 and the prior in \eqref{eq:prior_rho} as $\rho \sim Beta(1,1)$.
For the HDP, the total mass parameters $\alpha$ and $\beta$ are fixed and equal to 1.
For our model, we set the prior hyperparameters for the marginal prior \eqref{eq:prior_Sigma} of $\Sigma$ as $\nu = 100$ and $V = \mathbb{I}$  for all the simulated examples. 
For the CK-SSM, we followed the hyperparameter tuning outlined in their paper, except for
the parameters $a$ and $b$ 
that we fix to $a = 0.1$ and $b=0.5$.

As metrics to compare the density estimates, i.e. the posterior mean of the density
evaluated on a fixed grid, we use the Kullback-Leibler divergence and the Hellinger distance 
between the estimated density and the true one.

\subsection{Non-Gaussian simulated data}
\label{sec:nongaussian}
We consider three scenarios.
In each scenario we generate, for $I=6$ different areas, an i.i.d. sample from a density that is not  Gaussian: namely t-student ($t$), skew-normal ($SN$), chi-squared ($\rchi^2$) and Cauchy ($Ca$).   The matrix $G$ is fixed and represents a graph with only three connected components
%, namely areas 1 and 2, areas 3 and 4, and finally areas 5 and 6 
$\{1,2 \}$, $\{3,4 \}$, $\{5,6\}$.
The three scenarios differ in the number of data in each area and  in the data generating densities, as reported in Table~\ref{tab:sim_01_data}:
$t(\nu, \mu, \sigma)$ denotes the Student's $t$ distribution with $\nu$ degrees
of freedom, centered in $\mu$ and scaled by a factor $\sigma$; 
$SN(\xi, \omega, \alpha)$ 
denotes the Skew normal distribution with mean 
$\xi + \omega \alpha / \sqrt{1 + \alpha^2} \sqrt{2 / \pi}$,
$\rchi^2(k, 0, 1)$ denotes the standard chi-squared distribution with $k$ degrees of 
freedom and $Ca(0,1)$ the Cauchy distribution.
\begin{table}[t]
    \centering
    \resizebox{\textwidth}{!}{%
    \begin{tabular}{c|c|c|c|c|c|c|c}
          & Area & 1 &  2 &  3 &  4 &  5 &  6 \\
         \hline
         Scenario I & Density & $t(6, -4, 1)$ & $t(6, -4, 1)$ & $SN(4, 4, 1)$ & $SN(4, 4, 1)$ & $\rchi^2(3, 0, 1)$ & $\rchi^2(3, 0, 1)$\\
& $N_i$ & $1000$ & $1000$ & $1000$ & $1000$ & $1000$ & $1000$\\
\hline
Scenario II & Density & $t(6, -4, 1)$ & $t(6, -4, 1)$ & $SN(4, 4, 1)$ & $SN(4, 4, 1)$ & $\rchi^2(3, 0, 1)$ & $\rchi^2(3, 0, 1)$\\
& $N_i$ & $1000$ & $10$ & $1000$ & $10$ & $1000$ & $10$\\
\hline
Scenario III & Density & $t(6, -4, 1)$ & $t(6, -4, 1)$ & $SN(4, 4, 1)$ & $SN(4, 4, 1)$ & $Ca(0, 1)$ & $Ca(0, 1)$\\
& $N_i$ & $100$ & $100$ & $100$ & $100$ & $100$ & $100$\\
    \end{tabular}}
    \caption{Non-Gaussian simulated data: true densities and sample sizes for each area under all scenarios}
    \label{tab:sim_01_data}
\end{table}
They cover extremely different cases: in Scenario I a large number of data is available in each area, so that borrowing strength from nearby areas would be superfluous; we actually expect our model to perform worse than the HDP-mixture, being the latter fully nonparametric.
On the other hand, in Scenario II  there are three areas (2, 4 and 6) with few data points (only 10).
In this case, we expect our model to express its strength and give a better density estimate than the HDP-mixture, especially in those areas where few data are present.
Finally, Scenario III is an in-between condition, where not so many observations as in Scenario I are available in each area.
We also compare the results obtained with the CK-SSM mixture.

\begin{figure}[t]
    \centering
    \includegraphics[width=\linewidth]{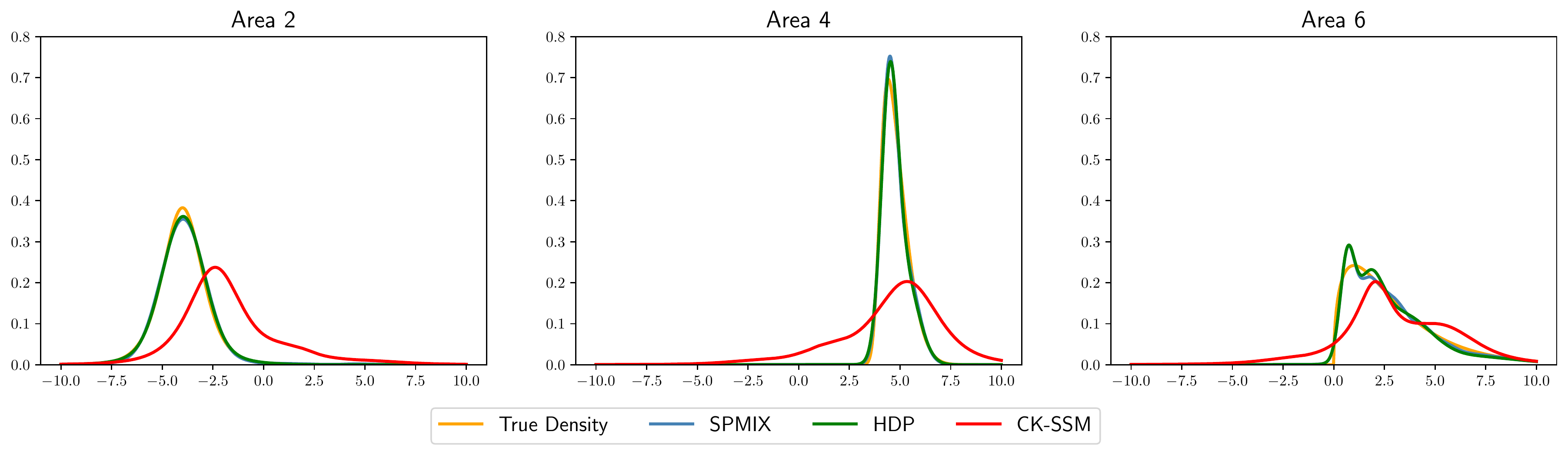}
\caption{Non-Gaussian simulated data, Scenario I: true densities 
for areas 2, 4 and 6 and the corresponding density estimates under the three different models.}
\label{fig:scen_1}
\end{figure}

\begin{figure}[t]
    \centering
    \includegraphics[width=\linewidth]{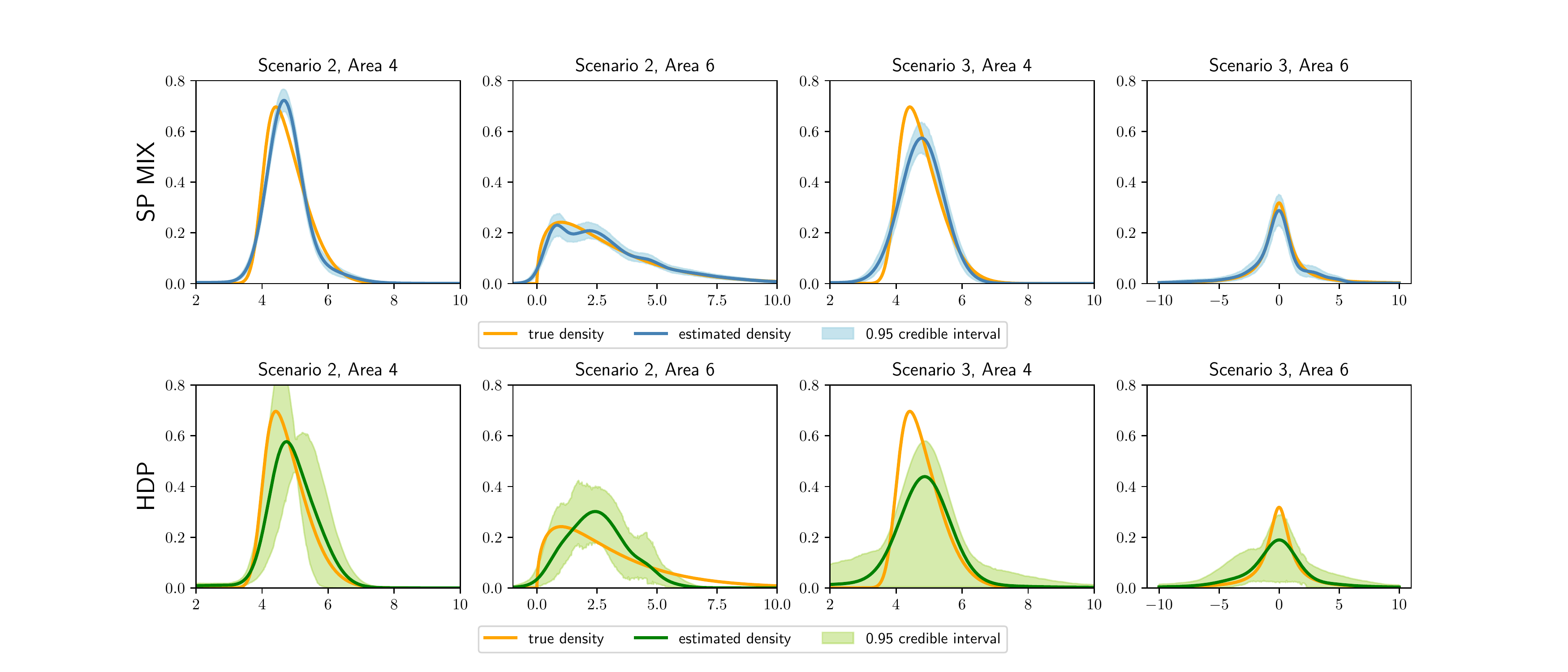}
    \caption{Non-Gaussian simulated data, Scenario II and III: estimated and true densities for areas 4 and 6 under our (top row) and HDP mixture (bottom row) models.}
    \label{fig:sim1}
\end{figure}

In order to make a fair comparison between our model, CK-SSM and the HDP models, we fixed the number of components $H$ in our mixtures and in the CK-SSM to 10.
This choice was made by looking at the posterior distribution of the number of components under the HDP-mixture in the different scenarios; we found that the
number of  clusters ranges between 3 and 10.
\begin{table}
    \centering
    \resizebox{\textwidth}{!}{%
    \begin{tabular}{c|c|c|c|c|c|c|c}
          & Model & 1 &  2 &  3 &  4 &  5 &  6 \\
         \hline
         Scenario I & SPMIX & $0.01 \pm 0.00$ & $0.01 \pm 0.00$ & $0.01 \pm 0.00$ & $0.01 \pm 0.00$ & $0.02 \pm 0.01$ & $0.02 \pm 0.01$\\
& HDP & $0.00 \pm 0.00$ & $0.00 \pm 0.00$ & $0.01 \pm 0.00$ & $0.01 \pm 0.00$ & $0.02 \pm 0.01$ & $0.02 \pm 0.01$\\
& CK-SSM & $0.92 \pm 0.46$ & $0.92 \pm 0.46$ & $0.97 \pm 0.16$ & $0.98 \pm 0.16$ & $1.10 \pm 0.31$ & $1.10 \pm 0.31$\\
\hline
Scenario II & SPMIX & $0.02 \pm 0.00$ & $0.04 \pm 0.04$ & $0.02 \pm 0.01$ & $0.02 \pm 0.07$ & $0.03 \pm 0.01$ & $0.03 \pm 0.10$\\
& HDP & $0.01 \pm 0.00$ & $0.13 \pm 0.04$ & $0.03 \pm 0.01$ & $0.21 \pm 0.07$ & $0.03 \pm 0.01$ & $0.32 \pm 0.10$\\
& CK-SSM & $0.91 \pm 0.40$ & $0.90 \pm 0.40$ & $0.97 \pm 0.17$ & $0.97 \pm 0.17$ & $1.22 \pm 0.45$ & $1.23 \pm 0.44$\\
\hline
Scenario III & SPMIX & $0.15 \pm 0.19$ & $0.15 \pm 0.18$ & $0.09 \pm 0.25$ & $0.09 \pm 0.25$ & $0.06 \pm 0.12$ & $0.06 \pm 0.12$\\
& HDP & $0.16 \pm 0.19$ & $0.16 \pm 0.18$ & $0.26 \pm 0.25$ & $0.26 \pm 0.25$ & $0.13 \pm 0.12$ & $0.13 \pm 0.12$\\
& CK-SSM & $0.86 \pm 0.33$ & $0.86 \pm 0.34$ & $1.25 \pm 0.29$ & $1.25 \pm 0.29$ & $0.86 \pm 0.41$ & $0.86 \pm 0.42$\\
    \end{tabular}}
    \caption{Kullback-Leibler divergences between the true densities and the estimated ones, aggregated over $100$ simulated datasets with $\pm$ one standard deviation}
    \label{tab:results_kl}
\end{table}
For each scenario we repeatedly simulated 100 independent datasets.
Table~\ref{tab:results_kl} shows
%the Hellinger distance and 
the KL-divergence  between the true density and the estimate under the
three models. We average those values over the 100 simulated datasets, also considering $\pm$ one empirical standard deviation of the 100 values obtained. Table~\ref{tab:results_hell} in Appendix~\ref{sec:app_additionalplots}, reports the same values for the Hellinger distance. 
From both tables, we can see that in all the three scenarios, the CK-SSM has 
the worst performance in recovering the true data generating density. 
This reflects what we discussed in Section~\ref{sec:comparison}: the prior of such model forces mixture weights to be too similar across different connected components in the graph.
We can clearly see this  for example from Figure~\ref{fig:scen_1}, where the density estimates for areas 4 and 6 (not connected in $G$) are close under the CK-SSM but not  under our model.
The HDP-mixture gives overall better estimates than those under our model in Scenario I. 
In this case, both density estimates  are close enough to the true densities; see Figure~\ref{fig:scen_1}.
As we expect, under Scenario II, our model gives a better density estimate (than the HDP-mixture) in areas 2, 4 and 6, where only 10 data points are available; see Figure~\ref{fig:sim1}. Indeed, our model  retrieves information from the neighboring areas, overcoming the lack of data in some of the areas.
Interestingly, our model performs better in areas 3-6, and similarly in areas 1 and 2, under Scenario III, probably because of 
\virgolette{extreme} data in areas 5 and 6, where we generate data from a Cauchy distribution.
This behavior is evident from Figure~\ref{fig:sim1}, being the $95\%$ point-wise credible interval of the posterior distribution of the density much wider in HDP than in our approach.
Finally,  it is clear that our model fits data well also when the true density  is highly non-symmetric, such as in locations 3-6 in Scenarios I and II.

%\bale Sono curiosa di vedere la posterior di $\rho$ \eale
%\mario{Ho messo un traceplot nel supplementary!}
%\bale Vista! Siccome \'e bruttina, meglio fare finta di niente e toglierla\eale

\subsection{Simulation from spatially dependent weights}\label{sec:spatial_weights}
\begin{figure}[h]
	\centering
	\hspace{-10mm}
	\begin{subfigure}[c]{0.3\textwidth}
		\centering
		\includegraphics[width=3.5cm]{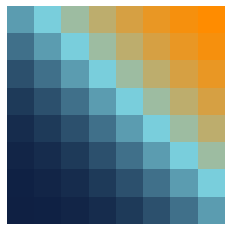}
		\caption{}
		\label{fig:first_weights_linear}
	\end{subfigure}
    \hspace{-7mm}
	\begin{subfigure}[c]{0.3\textwidth}
    	\centering
    	\includegraphics[width=3.5cm]{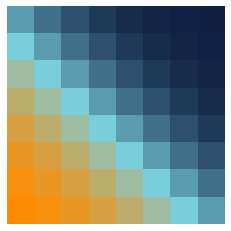}
    	\caption{}
    	\label{fig:second_weights_linear}
    \end{subfigure}
	\hspace{-5mm}
	\begin{subfigure}[c]{0.34\textwidth}
		\centering
		\includegraphics[width=5.5cm]{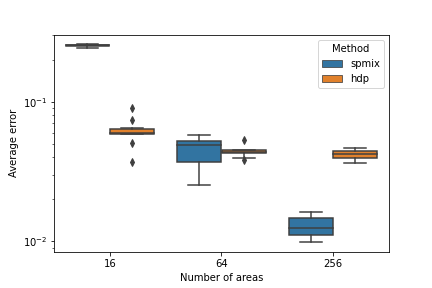}
		\caption{}
		\label{fig:kl_errors_sim_2}
	\end{subfigure}
\vspace{-3mm}
\caption{Simulation from spatially dependent weights in Sect. \ref{sec:spatial_weights}: plots of $\{w_{i1}\}$ (a), $\{w_{i2}\}$ (b) when $I=64$; boxplots (c) of the Kullback-Leibler divergence between true density \eqref{eq:spatial_truedens} and estimated one under our model (spmix) and the HDP-mixture model (hdp) for each simulation, averaged over the areas, for $I=16, 64, 256$, in logarithmic scale.}
\label{fig:loc_weights_linear}
\end{figure}

In the second simulation example we apply our model to estimate spatially dependent densities in contiguous areas, placed in a squared grid with a total number $I$ of areas, in a unit area squared domain; we study three different scenarios, choosing $I = 16, 64, 256$.
In the $i$-th area, we simulate observations as follows:
\begin{equation}
y_{ij} \iid  w_{i1} \calN(-5, 1) +  w_{i2} \calN(0, 1) + w_{i3} \calN(5, 1)\quad j=1, \ldots, 25
 \label{eq:spatial_truedens}
\end{equation}
where the weights are chosen as $alr^{-1}(\tilde{\bm w_i})$ and the transformed weights $\tilde{\bm w_i}$ are given by
\begin{equation} \label{eq:spatial_weights}
    \tilde{w}_{i1} = 3 ( x_i - \bar{x}) + 3 (y_i - \bar{y}) \quad  \tilde{w}_{i2} = -3 ( x_i - \bar{x}) - 3 (y_i - \bar{y})
\end{equation}
where $(x_i, y_i)$ are the coordinates of the center of area $i$ and $(\bar{x}, \bar{y})$ the coordinates of the grid center.
In this way, we introduce strong spatial dependence, induced by (\ref{eq:spatial_weights}), among the weights of different areas, as we observe in Figures~\ref{fig:first_weights_linear} and \ref{fig:second_weights_linear}, where we plot the weights of the first two components $\{w_{i1}\}$ and $\{w_{i2}\}$, for the scenario $I=64$; of course $w_{i3} = 1 - w_{i1} - w_{i2}$.

In our model, we consider areas $i$ and $j$ to be neighbors if they share an edge, setting $g_{ij} = 1$ in \eqref{eq:prior_weights} in this case, and $g_{ij} = 0$ otherwise.
For each scenario, we simulated 10 independent datasets, sampling 25 observations
per area, and then we compare the posterior estimates of the densities with the true
ones  via  Kullback-Leibler divergence. 
We compare our model with the HDP-mixture model, reporting in 
Figure~\ref{fig:kl_errors_sim_2} the errors, averaged over all areas, for the
ten repetitions. 
Though when $I=16$ HPD gives much better estimates, our model outperforms
the HDP-mixture, when the number of areas is sufficiently large, with consistent results using the Hellinger distance to measure the errors, as shown in Figure~\ref{fig:error_sim2_Hellinger} in Appendix~\ref{sec:app_additionalplots}.

The median execution times, over the 10 datasets, of the code corresponding to our model  was 25.28, 118.14 and 616.41 seconds  for $I=16, 64, 256$, respectively, 
whereas, for fitting data for the HPD-mixtures  was 18.39, 72.59, 207.46 seconds. 
Based on our implementation, HDP is slightly faster, but our model still exhibits competitive computational times, paired with lower errors when the number of areas is large.
Simulations were performed on a machine equipped with a 4x Xeon E5-2640 v4 @2.4GHz processor and 64 GB of RAM.
In order to provide a fair comparison, the implementation for
our model ran on a single core since the sampler for the HDP is inherently sequential.
However, the Gibbs sampler we proposed can be straightforwardly parallelized and this
could greatly decrease the runtimes of our model.

%\rev{1}{The computing times are noted, but MCMC efficiency (e.g., effective sample sizes)
%are not discussed. Since the Gibbs updates are not blocked across h—the steps are
%componentwise $[ \wtilde_{ih} | \wtilde_{i,-h}, \cdots]$—efficiency is a concern. It would be more appropriate to report computing time per effective sample (e.g., seconds per 1000 effective samples) for key parameters of interest.}
%\mario{Qua confrontiamo semplicemente HDP e il nostro modello, non abbiamo parametri comuni su cui confrontare ESS quindi lascerei perdere.}

\section{Airbnb Amsterdam}
\label{sec:airbnb}

%\rev{1}{What is learned (or improved upon) by using the proposed mixture model, e.g.,
%rather than a basic Gaussian linear regression model? The case for the methodology
%is not totally convincing from this application.}

%\rev{2}{Comparison with a mixture model where the weights of the mixture do not vary with the area (?) but the parameters of the Gaussian components vary spatially - see b) and c) - Questo dipende un po' dal refree che abbiamo, forse si pu\`{o} lasciare perdere per il momento? O possiamo forzare il nostro modello a non avere la dipendenza spaziale? }

%\rev{2}{In my opinion, the data analysis Section is lacking in several ways. The modeling choices are not completely clear (the number of accommodates is clearly positively correlated to the number of bedrooms - why the decision to add both?). More importantly, why is it important to predict the differential pricing in each neighborhood? Of course, a landlord may be interested in finding out how to price their property, but this (or similar) justification is never mentioned. Nevertheless, why is density estimation important, more than just predicting an average price? Isn’t a booking price all that matters to a booking agency? Why should a landlord be interested in the tails? Note: one can come up with many reasons for looking at densities; I do not deny the method's utility (far from that, actually). I am just saying that a reader is left to impute a lot of missing data here.}

We consider the Airbnb listings dataset for the city of Amsterdam (The Netherlands), publicly available at \url{http://insideairbnb.com/get-the-data.html}.
The dataset consists of more than $20,000$ listings spread over Amsterdam,
grouped by the neighborhood.
Our main goal is the prediction of the nightly price of a new listing, with information given by covariates, and 
taking into account the spatial dependence.  
% As already stated in the Introduction, we believe that density modeling within this context could be really valuable, especially when landlords need to take some decision about the positioning of their flats on the market. In fact, the posterior density estimation carries a much richer information than the point estimate of the average price, leading to better informed decisions about their market strategy.}
As mentioned in the Introduction, (joint) density modeling and estimation, in this case, can give insight to landlords who need to take decision on where their flats should be positioned in the flat rental market. In fact, full posterior density estimate carries much richer information than a simple point estimate of the average price, leading to better informed decisions about the market strategy.
We consider two generalizations of model \eqref{eq:lik} to account for covariates as follows.
Denote responses as $y_{ij}$ (i.e. the nightly price of accommodation $j$ in neighborhood $i$) and covariates as $\bm x_{ij}=(x_{ij1},\ldots,x_{ijd})^T$. In the first model, denoted here $M1$,  we assume $\tau_h$ in  \eqref{eq:lik} such that 
$\tau_h=(\mu_h+\bm \beta^T \bm x_{ij}, \sigma^2_h)$, $h=1,\ldots,H$.
$M1$ can be understood as a linear regression model with component-specific intercept and variance. 
We further generalize $M1$ by assuming that all the regression coefficients are component-specific, i.e. $\tau_h=(\mu_h+\bm \beta_h^T \bm x_{ij}, \sigma^2_h)$, $h=1,\ldots,H$, and denote it by $M2$.
While model $M1$ assumes that the effect of the covariates on the pricing is shared across all neighborhoods, and the spatial effect can be represented by the only intercept, model $M2$ assumes that all the covariates have different effect on the pricing depending on the neighborhood.

%For this reason, we generalize the model adding one more level to the likelihood (a regression model). Specifically, we assume  for responses $y_{ij}$ (i.e. the nightly price of accommodation $j$ in neighborhood $i$), conditional to $d$ covariates $\bm x_{ij}=(x_{ij1},\ldots,x_{ijd})^T$ and parameters, the same model \eqref{eq:lik}-\eqref{eq:prior_rho}, where $\tau_h$ in  \eqref{eq:lik} are such that 
%$\tau_h=(\mu_h+\bm \beta^T \bm x_{ij}, \sigma^2_h)$, $h=1,\ldots,H$.
%The prior specification is completed assuming 
%\begin{align}
%%    y_{ij} &= \bm \beta^T \bm x_{ij} + \varepsilon_{ij} \label{eq:regression} \\
%    \bm \beta & \sim \calN_d(\bm 0, \sigma^2_{\beta} \mathbb{I}_d) \label{eq:prior_beta}
%\end{align}
%where $\bm \beta=(\beta_1,\ldots,\beta_d)^T$ is the vector of regression parameters common to all the areas.
% and the residuals $\{\varepsilon_{ij}\}$ are distributed according to our model \eqref{eq:lik} - \eqref{eq:prior_rho}.

%\AGsub{In Section~\ref{sec:exploratory} we report a detailed description and exploratory analysis of the dataset.}{}

\subsection{Data description}
%%\subsection{Data description and exploratory data analysis}
\label{sec:exploratory}
We consider as predictors characteristics of the house such as:
\begin{enumerate*}[label=(\roman*)]
    \item \texttt{accommodates}, the number of guests that can be hosted,
    \item \texttt{bathrooms}, the number of bathrooms,
    \item \texttt{bedrooms}, the number of bedrooms;
\end{enumerate*}
together with two indicators of popularity of the listing:
\begin{enumerate*}[label=(\roman*)]
   \setcounter{enumi}{3}
   \item \texttt{number\_of\_reviews}, the number of reviews present for that listing, and
   \item \texttt{review\_scores\_rating} the average rating of the reviews.
\end{enumerate*}
Finally, we complete the set of covariates with two binary variables:
\begin{enumerate*}[label=(\roman*)]
   \setcounter{enumi}{5}
   \item \texttt{instant\_bookable} which equals $1$ if the listing can be booked instantly from the user and $0$ if, instead, the request must go through an acceptance procedure from the host;
   \item \texttt{host\_is\_superhost} that is $1$ if the host is classified as a \emph{superhost} by Airbnb and 0 otherwise. The \emph{superhost} badge can be obtained once a host has a sufficient number of reviews with a rating above a certain threshold.
\end{enumerate*}
These binary  variables were included since the user, while searching for an accommodation, can reduce her/his search only to instant bookable listings and/or only to \emph{superhosts}.

As preprocessing, we used the following steps: we removed the listings for which at least one predictor is missing,  as well as listings whose nightly price is below two euros or above one thousand euros; then we transformed \texttt{number\_of\_reviews} by taking the natural logarithm and 
\texttt{review\_scores\_rating} by the Box-Cox transformation 
$x_i^{(\lambda)} =  (x_i^\lambda - 1)/\lambda$ \citep{box1964analysis}
with  $\lambda=12$, being this value automatically chosen by the Python package \texttt{scipy}.

\begin{figure}[t]
	\centering
	\begin{subfigure}{0.33\textwidth}
		\centering
		\includegraphics[width=\linewidth]{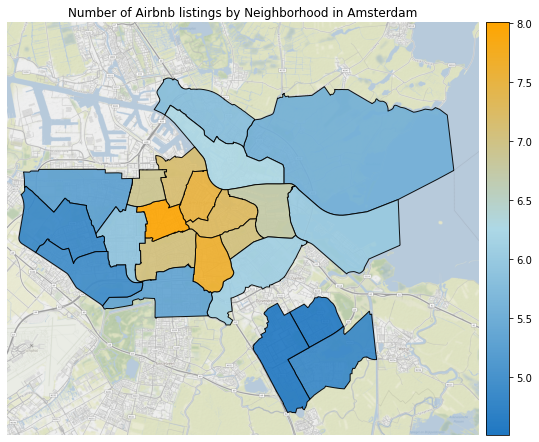}
		\caption{No. of listings in the log-scale}
	\end{subfigure}%
	~
	\begin{subfigure}{0.33\textwidth}
		\centering
		\includegraphics[width=\linewidth]{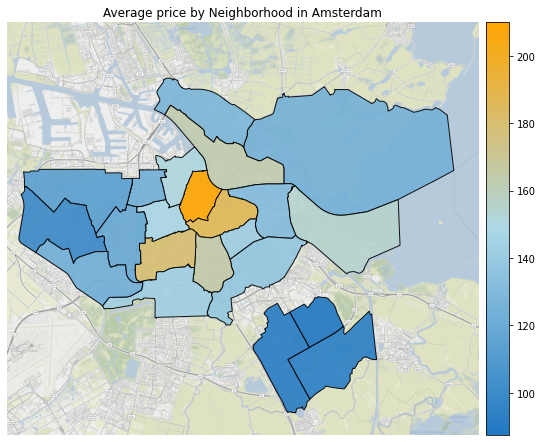}
		\caption{Sample mean}
	\end{subfigure}%
	~
	\begin{subfigure}{0.33\textwidth}
		\centering
		\includegraphics[width=\linewidth]{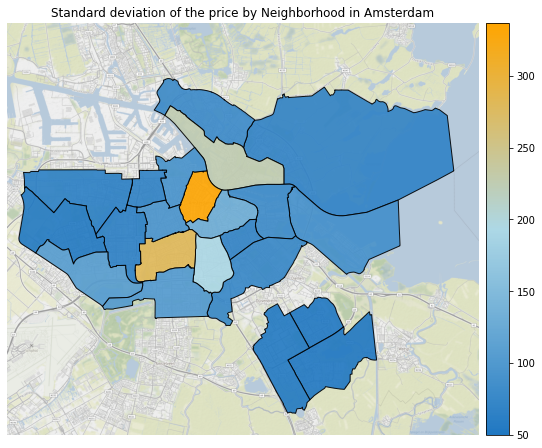}
		\caption{Sample standard deviation}
	\end{subfigure}
	\caption{Number of listings (in the log-scale), sample means and standard deviations of the nightly price  in euros for each neighborhood in the Airbnb dataset, after preprocessing.}
	\label{fig:nlist_mean_std_by_neigh}
\end{figure}

%%\begin{figure}[h!]
%%	\centering
%%	\includegraphics[width=0.5\linewidth]{images/count_by_neigh}
%%	\vspace*{-1mm}
%%	\caption{Number of listings for each neighborhood in the Airbnb dataset, after preprocessing, in the log-scale.}
%%	\label{fig:count_by_neigh}
%%\end{figure}
%\subsection{Exploratory data analysis}\label{sec:exploratory}
Each listing is assigned to one of the twenty-two Amsterdam neighborhoods provided at the dataset web page, so that $I=22$. The total number of observations considered for our analysis is $N_1+\cdots+N_I=17,201$.
Figure~\ref{fig:nlist_mean_std_by_neigh}(a) shows  sample sizes in the log-scale for each neighborhood; of course, city center is the area with the largest number of observations. Furthermore, in 
Figure~\ref{fig:nlist_mean_std_by_neigh}, panels (b) and (c), we report sample means and standard deviations of the nightly price in euros in each neighborhood; the plots motivate the modeling of the spatial dependence, as close neighborhoods tend to have similar distributions, at least in terms of mean and standard deviation.  
Figure~\ref{fig:nlist_mean_std_by_neigh} shows that there are two distinct graph connected components, one made only by three areal units; this agrees with official neighborhood maps of the city of Amsterdam.
As far as covariates are concerned, Figure~\ref{fig:corr_matrix} in Appendix~\ref{sec:app_additionalplots} shows empirical correlations  among the predictors and between predictors and the response.  Figure~\ref{fig:predictors_plots} in Appendix~\ref{sec:app_additionalplots} displays  scatterplots of the response price versus numerical predictors and boxplots for categorical predictors. We note that only \texttt{accommodates}, \texttt{bathrooms}, \texttt{bedrooms} exhibit a significant linear correlation with the price, which is confirmed by the scatter plots, while sample linear correlation between \texttt{accommodates} and \texttt{bathrooms} is 0.362, 0.730 between \texttt{accommodates} and \texttt{bedrooms}, 0.430 between \texttt{bathrooms} and \texttt{bedrooms}. However, when computing the variance inflation factor, we found 2.197, 1.243 and 2.334, respectively, values that suggest very mild multicollinearity. On the other hand,  there is no evident empirical effect of \texttt{instant\_bookable} and \texttt{host\_is\_superhost} on the nightly price as Figure~\ref{fig:predictors_plots} in Appendix~\ref{sec:app_additionalplots} shows.   
In the next subsection, we consider both models $M1$ and $M2$ for the dataset, including all the covariates above described, i.e. $d=7$.  We standardized all numerical predictors, subtracting the sample mean and dividing by the sample standard deviation of each predictor; we also centered the response on the overall sample mean.
 
\subsection{Posterior inference}
 
We complete the prior for   model $M1$ assuming 
$$(\mu_h, \sigma^2_h) \iid \calN(\mu_h \mid 0, 2\sigma_h^2 )\times IG(\sigma^2_h \mid 2, 2), \quad h=1,\ldots,H$$ and  
$\bm \beta \sim \calN_d(\bm 0, \sigma^2_{\beta} \mathbb{I}_d)$. 
For the prior of model $M2$ we assume
%(\widetilde{\bm\beta_h}
$$( (\mu_h,{\bm\beta_h}), \sigma^2_h) \iid \calN((\mu_h,{\bm\beta_h}) \mid \bm 0, 10 \mathbb{I}_{d+1}) \times IG(\sigma^2_h \mid 2, 2), \quad h=1,\ldots,H.$$
We need to change the Gibbs sampler in Section~\ref{sec:gibbs}, adding two further steps  to  sample from the full conditional of $\bm\beta$ for model $M1$ or  from the full conditional of $(\mu_h, \beta_h)$ for model $M2$. Both steps are standard updates in Bayesian linear regression models; see Appendix~\ref{sec:app_Gibbs} for further details. 

Posterior inference is robust to the choice of all the
hyper-parameters in the prior distribution, but for the number $H$ of components in the mixture, that is a key parameter for mixture models. 
For this reason, we compare $M1$ and $M2$ via predictive goodness-of-fit indexes such as the log-pseudo marginal likelihood \citep[LPML,][]{geisser1979predictive} and the  widely applicable information criterion
 \citep[WAIC,][]{watanabe2013widely}, when $H$ varies in $\{5, 10, 15\}$.  Better predictive performances are associated to higher LPML and lower WAIC.
 In this comparison, we also consider a generalization of the CK-SSM model in \cite{jo2017dependent} along the lines of model $M1$.
Table~\ref{tab:amsterdam} shows that the  best model is $M1$, across all values of $H$ and that CK-SSM does a worse job than $M1$ and $M2$. 
 Given its superior predictive performance, in the following we consider only $M1$.

In particular, $M1$ with $H=15$ gives slightly better values of LPML and WAIC, but the difference across all values of $H$ seems negligible so that,
to fix $H$, we consider also the predictive mean squared error computed through a 10-fold cross-validation.
The cross-validation is stratified according to the areas, so that, each time, approximately $10\%$ of the data is missing from each neighborhood.
More in detail, each time we select $90\%$ of the dataset as \virgolette{training set} (to simulate from  the relative posterior) and compute the mean of the predictive distribution corresponding to data in the \virgolette{testing set}. Observe that the same datapoints are shared across all values of $H$, both for training and for testing. Then we compute the predictive mean squared error (pMSE) on the testing set,
i.e.   $\sum_{i=1}^m (y_i - \hat{y}_i)^2 / m$, where $m$ is the size of the testing set and $\hat{y}_i$ is the mean of the posterior predictive density of the response corresponding to covariate $\bm x_i$. 
The average cross validation error $\pm$ one standard deviation is equal to 5468 $\pm$ 952, 5474 $\pm$ 850 and 5477 $\pm$ 956 for $H=5, 10, 15$ respectively.

We have also considered the case $H=1$, i.e. when  all $M1$, $M2$ and CK-SSM models are equivalent to a standard Gaussian linear regression. In this case, the predictive performance is much worse (LMPL and WAIC are approximately equal to $-1.3 \times 10^5$ and $2.6 \times 10^5$ respectively), hence showing that a richer model with explicit modeling of the spatial dependence structure is needed to obtain better predictive performances.
Moreover, removing covariates \texttt{bathrooms} and \texttt{bedrooms},  correlated with \texttt{accommodates}, resulted in slightly worse predictive performance for all models tested; for instance,  $M1$ showed a decrease in LMPL of $2.5\%$, while for $M2$ the decrease was around $1\%$ across all values of $H$.  
Summing up, for the reasons above, including parsimony of the model, in the rest of the section, we consider only model $M1$ when $H=5$.

\begin{table}
\centering
\resizebox{\textwidth}{!}{%
\begin{tabular}{c|c c c | c c c | c c c}
& \multicolumn{3}{c|}{$M1$} & \multicolumn{3}{c}{$M2$} & \multicolumn{3}{c}{CK-SSM} \\
$H$ & $5$ & $10$ & $15$ & $5$ & $10$ & $15$ & $5$ & $10$ & $15$ \\ \hline
LPML & -92619 & -92444 & -92441 & -97998 & -97836 & -97828 & -98751 & -98752  & -98755 \\
WAIC & 185238 & 184888 & 184882 & 195996 & 195672 & 195656 & 197502 & 197504 & 197505 \\
%pMSE & 5468 $\pm$ 952 & 5474 $\pm$ 850 & 5477 $\pm$ 956  & $-$ & $-$ & $-$
\end{tabular}
}
\caption{LPML and WAIC for various choices of $H$ under $M1$, $M2$ and CK-SSM.}
\label{tab:amsterdam}
\end{table}

\begin{figure}[h!]
    \centering
    \begin{subfigure}{0.4\textwidth}
        \centering
        \includegraphics[width=\linewidth]{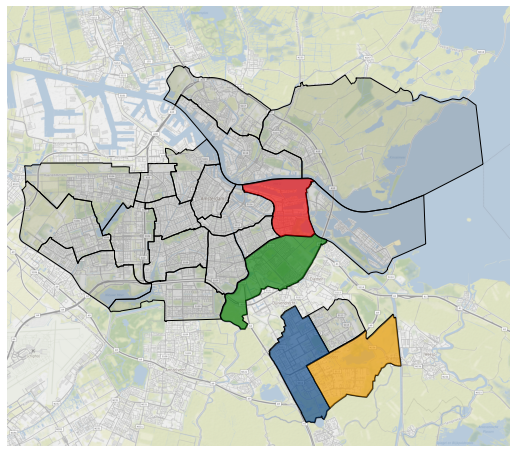}
        \caption{}
 \end{subfigure}\hfill
   \begin{subfigure}{0.45\textwidth}
        \centering
        \includegraphics[width=0.8\linewidth]{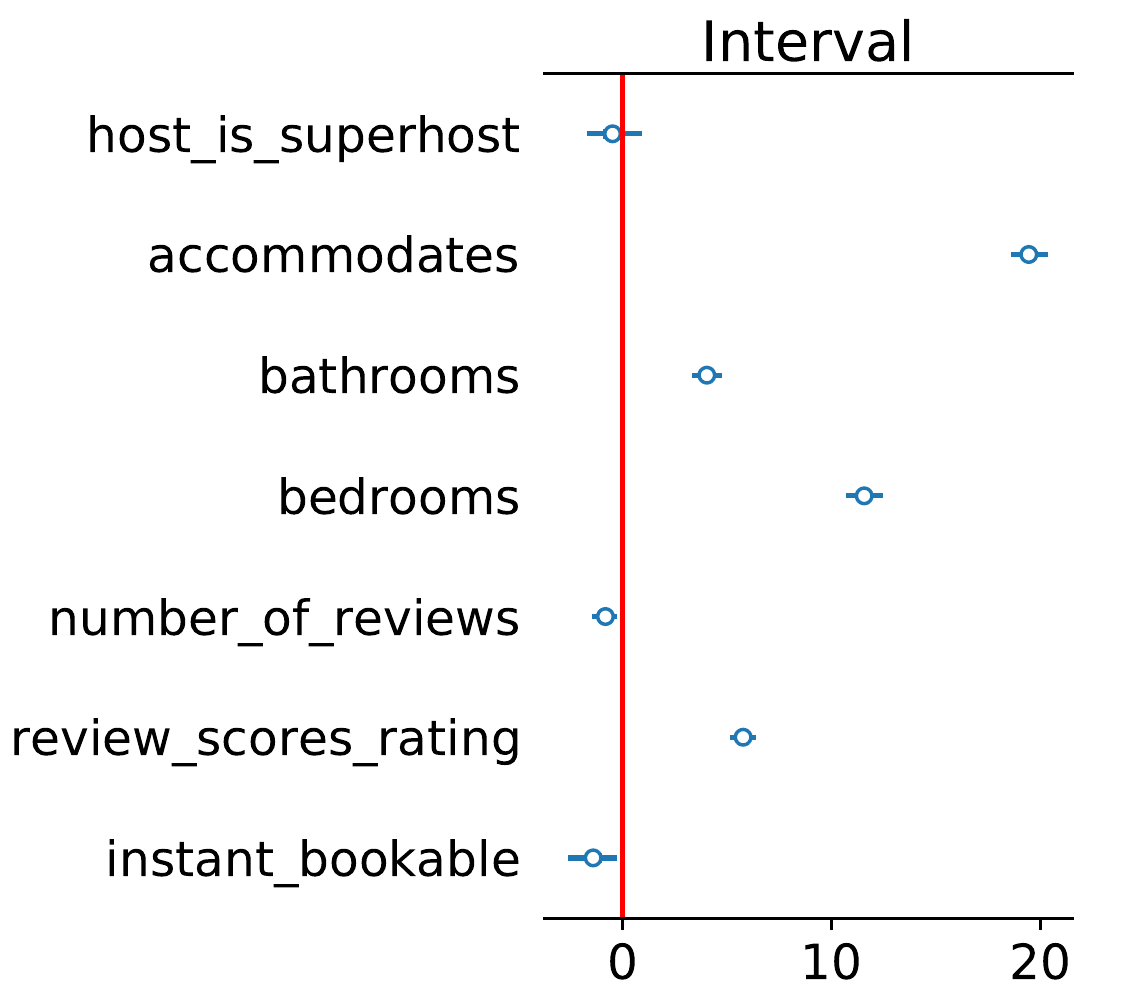}
        \caption{}
    \end{subfigure}
   
   \begin{subfigure}{0.8\textwidth}
        \centering
        \includegraphics[width=\linewidth]{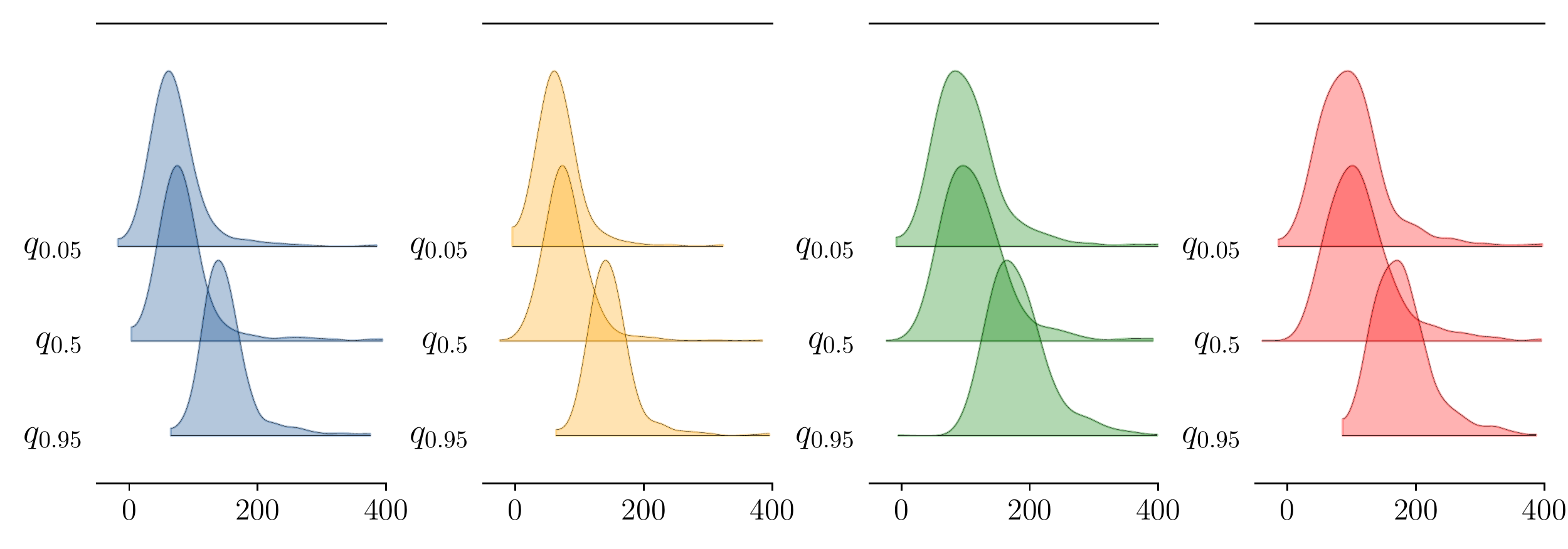}
        \caption{}
    \end{subfigure}
	\vspace{-4mm}
    \caption{
    (a): Map of the city of Amsterdam with  neighborhoods \emph{Bijlmer-Centrum} in blue, \emph{Gaasperdam - Driemond} in orange, \emph{Oostelijk Havengebied - Indische Buurt} in green and \emph{Watergraafsmeer} in red. 
    (b): 95\% credibility intervals of the marginal posterior of the regression parameter $\bm \beta$.
    (c): Predictive densities for different neighborhoods, the colors match the ones of
     the map.
        In each plot, three lines represent three different values of the covariates 
        \texttt{accommodates}, \texttt{number\_of\_bedrooms}, \texttt{number\_of\_bathrooms}) that take range in the first, second and third quantile, while the other
        numerical covariates are fixed to the empirical median.
}
    \label{fig:res_dens}
\end{figure}

Figure~\ref{fig:res_dens}(b) reports $95\%$ posterior credibility intervals for the regression parameters.
All the covariates, except for \texttt{host\_is\_superhost}, seem to be significant, if we assume hard shrinkage as a criterion for significance, i.e. the marginal credibility interval does not include 0.
It is interesting to observe how the coefficients associated to  \texttt{number\_of\_reviews} and \texttt{instant\_bookable} are negative.
This might indicate that hosts that receive many reviews and many reservations tend to lower their prices in order be more attractive.
On the other hand, as one would expect, all the other coefficients are positive, the one furthest right being associated to \texttt{accommodates}, i.e. the number of guests that can be hosted.
%
%Figure \ref{fig:res_dens} shows the predictive distribution in different settings. 
In Figure~\ref{fig:res_dens}(c) we show the density estimates
in the four neighborhoods highlighted in Figure~\ref{fig:res_dens}(a).
Each plot shows three density estimates, corresponding to 
different values of the covariates.
In this case,  the covariates were set to the empirical median except for \texttt{accommodates}, \texttt{number\_of\_bedrooms}, \texttt{number\_of\_bathrooms}.
Since the marginal sample correlation between these three covariates is not negligible as mentioned in Section~\ref{sec:exploratory},
% (see Figure~\ref{fig:corr_matrix} in Appendiix\ref{sec:app_additionalplots}), 
we have fixed all their values simultaneously equal to 5\%, 50\% and 95\%  empirical quantiles, respectively. 
For instance, 
in each panel of Figure~\ref{fig:res_dens}(c), the top lines correspond to density estimates for a vector of covariates in which  \texttt{accommodates}, \texttt{number\_of\_bedrooms}, \texttt{number\_of\_bathrooms} are fixed to their  5\% sample quantile, respectively.
It is clear from Figure~\ref{fig:res_dens}(c) that the predictive densities in blue (first panel from the left) and in yellow are similar, as well as the lines  in green and in red.  
However there are evident differences when comparing for instance the yellow densities (second panel from the left) with the green ones (third panel from the left); indeed the green densities give substantial mass to the right tail, especially to values greater than $200$ euros, while the yellow densities do not.
%(\bale Perch\'{e} non calcoliamo la prob, sotto tutte queste estimates, che il prezzo sia $>200$? Cos\`{i} rispondiamo alla mia domanda in fondo a questa section \eale)
This behavior agrees with the marginal posterior of $\rho$, that is strongly concentrated near $1$ ($\E(\rho|data)=0.993$): in fact, 
the blue and yellow neighborhoods, as well as  the red and green ones, are connected in the graph. However,  blue and yellow predictive densities are  different from the green and red estimates, since the  neighborhoods belong to different connected components in the graph.
As expected, in all the neighborhoods the listings price increases 
as the \texttt{accommodates}, \texttt{number\_of\_bedrooms}, \texttt{number\_of\_bathrooms} increase as well.
Finally, for a new listing $y^\star$ with associated covariates $x^\star$, 
Table~\ref{tab:p200}, Appendix~\ref{sec:app_additionalplots}, reports the posterior predictive probability $P(y^\star > 200 \mid x^\star, i)$,
assuming that the new listing belongs to either one of the four neighborhoods and that the covariates $x^\star$ are equal to values as in  Figure~\ref{fig:res_dens}(c).

To conclude, we believe that a new lessor could benefit from our analysis because Airbnb makes available only a handful of covariates which might not be suited to fully characterize the \virgolette{right} nightly price for a listing. For instance,  we expect that the presence of a balcony or garden might lead to higher prices.
Hence, when deciding the price for a listing, the lessor could look at the predictive distribution from our model given the covariates and neighborhood of their house, and choose to place the listing in the right or left tail of the predictive distribution considering additional information not included in our model.

\section{Discussion}
\label{sec:discussion}
In this work, we have considered the problem of the joint estimation of 
spatially dependent densities in the context of repeated areal measurements. We have presented a finite mixture model to represent the density
in each area; assuming that all the mixtures share the same set of atoms,
the spatial dependency has been introduced through a novel joint distribution
for $I$  vectors in the simplex as a prior for the mixture weights.
This distribution, that we termed logisticMCAR, was built  as a
logistic  transformation of a specification of the multivariate CAR model.
When compared to alternatives proposed in the literature, the logisticMCAR
distribution showed to have a higher degree of interpretability, as we were
able to derive the analytic expression for the expected values of ratios of components and their covariances, via  the Aitchison geometry.
Moreover, we also showed as the
logisticMCAR can be used to accurately model sparse mixtures.

Posterior simulation  has been carried out by means of a Gibbs sampler  scheme.
In particular the update of the mixture weights was performed by introducing
a data augmentation scheme based on the P\'olya-Gamma identity, which avoids
the tedious tuning of the proposal distribution.

In the simulation studies and the real application  included in this paper,
our model has shown to be able to  represent a wide range of different behaviors.
In particular, we argue that when different connected graph components are present,
and heterogeneous behavior is observed across these components,
our model should be preferred as it does not force the densities in different graph component to behave too similarly. 
Moreover, as  in the case of the Airbnb Amsterdam application, our model can be easily extended
to include  additional covariate information.
Although not our target here, a sub-product of the approach is the prior induced on the partition of the subjects in the sample, which in this case, has a spatial connotation; relations with spatial product partition models \citep{page2016spatial} could be further investigated. 

Another point that we did not address here, and will be 
focus of future study,
is an extension to models where the \textit{graph} $G$ is not fixed, and should be learned by the data (and the prior). In particular, we aim at considering
boundary detection problems, i.e. when the proximity matrix $G$ is unknown, but its  elements depend on dissimilarity metrics
available for each pair of neighboring areas. This is an extremely interesting problem, widely studied in the context of one single response per area;  
see, for instance \cite{lu2007bayesian} and \cite{lee2012boundary}. However 
 preliminary investigation showed how the non-identifiability
of overfitted mixtures might produce erroneous results.
Possible extensions of our model to account for boundary detection might 
then include either a prior on the number of components or a repulsive prior distribution on the atoms, or both, to reduce the 
impact of non-identifiability.

\clearpage
\appendix

\section{Proofs}\label{sec:app_proofs} \hfill\\
\noindent
\underline{\textbf{Proof of Proposition \ref{prop:cond_means_s}}} \\
From equation \eqref{eq:mcar_tilde} we have that
\begin{align*}
\E \left[ \bm \wtilde_i \mid \bm \wtilde_{-i} \right] &= 
\frac{\rho \sum_{j \in U_i}\bm \wtilde_{j} + (1 - \rho) \bm \mtilde_i}{\rho |U_i| + 1 - \rho} = 
\frac{\rho \sum_{j \in U_i}alr(\bm w_{j}) + (1 - \rho) alr(\bm m_i)}{\rho |U_i| + 1 - \rho} \\&= 
\frac{1}{\rho |U_i| + 1 - \rho} \left(\log \frac{\prod_{j \in U_i} w_{j1}^\rho m_{i1}^{1-\rho}}{\prod_{j \in U_i} w_{jH}^\rho m_{iH}^{1-\rho}}, \ldots, \log \frac{\prod_{j \in U_i} w_{jH-1}^\rho m_{iH-1}^{1-\rho}}{\prod_{j \in U_i} w_{jH}^\rho m_{iH}^{1-\rho}} \right) \\
&= \frac{1}{\rho |U_i| + 1 - \rho} \left( \sum_{j \in U_i}  \log (w_{j1}^\rho m_{i1}^{1-\rho}), \ldots, \sum_{j \in U_i}  \log (w_{jH-1}^\rho m_{iH-1}^{1-\rho}) \right) \\&- \sum_{j \in U_i}  \log (w_{jH}^\rho m_{iH}^{1-\rho})
\end{align*}
where the last subtraction is meant elementwise.
Hence we have that
\begin{align*}
    \E \left[\log \frac{w_{il}}{w_{ik}}\mid \bm w_{-i} \right] &= 
    \E \left[ \wtilde_{il} - \wtilde_{ik} \mid \bm w_{-i} \right] \\
    &=\frac{1}{\rho \left| U_i \right| +1-\rho} \left( \sum_{j \in U_i}  \log (w_{jl}^\rho m_{il}^{1-\rho}) - \sum_{j \in U_i}  \log (w_{jk}^\rho m_{ik}^{1-\rho}) \right) \\
    &= \log \left(\left(\frac{m_{il}}{m_{ik}}\right)^{1-\rho} \prod_{j \in U_i} \left(\frac{w_{jl}}{w_{jk}}\right)^\rho \right)^{\frac{1}{\rho |U_i| + 1 - \rho}}
\end{align*}
which proves the proposition. 

\bigskip
\noindent
\underline{\textbf{Proof of Proposition \ref{prop:marg_cov}}} \\
The (marginal) joint distribution of two different components of $\bm \wtilde_i, \bm \wtilde_j$, with $ i,j=1,\ldots,I$, $i\neq j$ can be easily derived from \eqref{eq:joint_car}:
\[
    \begin{pmatrix} \wtilde_{il} \\ \wtilde_{jm} \end{pmatrix} \sim \calN_{2} \left(\bm 0, \begin{bmatrix} A_{ii}\Sigma_{ll} & A_{ij} \Sigma_{lm} \\A_{ji} \Sigma_{ml} & A_{jj}\Sigma_{mm} \end{bmatrix} \right)   \quad l,m=1,\dots,H-1
\]
Hence, we compute the covariance of the log ratios of different components as
\begin{align*}
  \Cov &\left(\log \frac{w_{il}}{w_{im}}, \log \frac{w_{jl}}{w_{jm}} \right)  = \text{Cov} \left(\wtilde_{il} - \wtilde_{im}, \ \wtilde_{jl} - \wtilde_{jm} \right)  \\
    &= \text{Cov} \left(\wtilde_{il}, \wtilde_{jl}\right) + \text{Cov} \left(\wtilde_{il}, \wtilde_{jm}\right) + 
    + \text{Cov} \left(\wtilde_{im}, \wtilde_{jl}\right) + \text{Cov} \left(\wtilde_{im}, \wtilde_{jm}\right) \\
    &= A_{ij}\left(\Sigma_{ll} - 2\Sigma_{lm} + \Sigma_{mm}\right)
\end{align*}
whereas, for the last component,
\begin{align*}
\Cov \left(\log \frac{w_{il}}{w_{iH}}, \log \frac{w_{jl}}{w_{jH}} \right) & = \text{Cov} \left(\wtilde_{il} , \ \wtilde_{jl} \right) = A_{ij}\Sigma_{ll} 
\end{align*}
which proves the formula in the proposition.  

It is possible to rearrange the indices $1, \ldots, I$ in order for $(F - \rho G)$ to be
a block diagonal matrix, where each block corresponds to a connected graph component according
to the neighboring structure; this will not  affect the joint law. 
By the properties of strictly diagonally dominated matrices, the same pattern of blocks is preserved in the inverse matrix $A$.
Hence $A_{ij} = 0$ if $i$ and $j$ belong to two non-connected graph components, proving the 
proposition. 

\section{MC simulations from the logisticMCAR distribution}
\label{sec:app_MC}  \hfill\\ \noindent
In Section~\ref{sec:logisticMCAR} we have pointed out that the theoretical analysis of the logisticMCAR distribution is limited
by its analytic intractability. 
Here we compute covariances between different components of the vectors of weights and Euclidean distances between the vectors themselves through Monte Carlo simulation. Specifically, we 
simulate from \eqref{eq:joint_car} and then obtain draws from the logisticMCAR distribution through the transformation $\alr^{-1}$.

In particular, we fix $I=5$, $H=3$, $\bm \mtilde_i = 0$ for all $i$ and the covariance matrix $\Sigma$
\[
\Sigma = 
    \begin{bmatrix} 
    1 & \Sigma_{12} \\
    \Sigma_{12} & 1 
    \end{bmatrix}
\]
where $\Sigma_{12}$ denotes the covariance, but also the correlation since $\Sigma_{11}=\Sigma_{22}=1$, between $\wtilde_{i1}$ and $\wtilde_{i2}$. We fix the proximity matrix $G$  such that $g_{12}=g_{13}=g_{23}=1$ and $g_{45}=1$. This corresponds assuming that areal units/nodes 1, 2 and 3 are connected to each other, and 4 and 5 are connected to each other, though separated from the others. 
%\[
%    G = 
%    \begin{bmatrix}
%    0 & 1 & 1 & 0 & 0 \\
%    1 & 0 & 1 & 0 & 0 \\
%    1 & 1 & 0 & 0 & 0 \\
%    0 & 0 & 0 & 0 & 1 \\
%    0 & 0 & 0 & 1 & 0
%    \end{bmatrix}
%\]
\begin{figure}[ht]
    \centering
    \includegraphics[width=\linewidth]{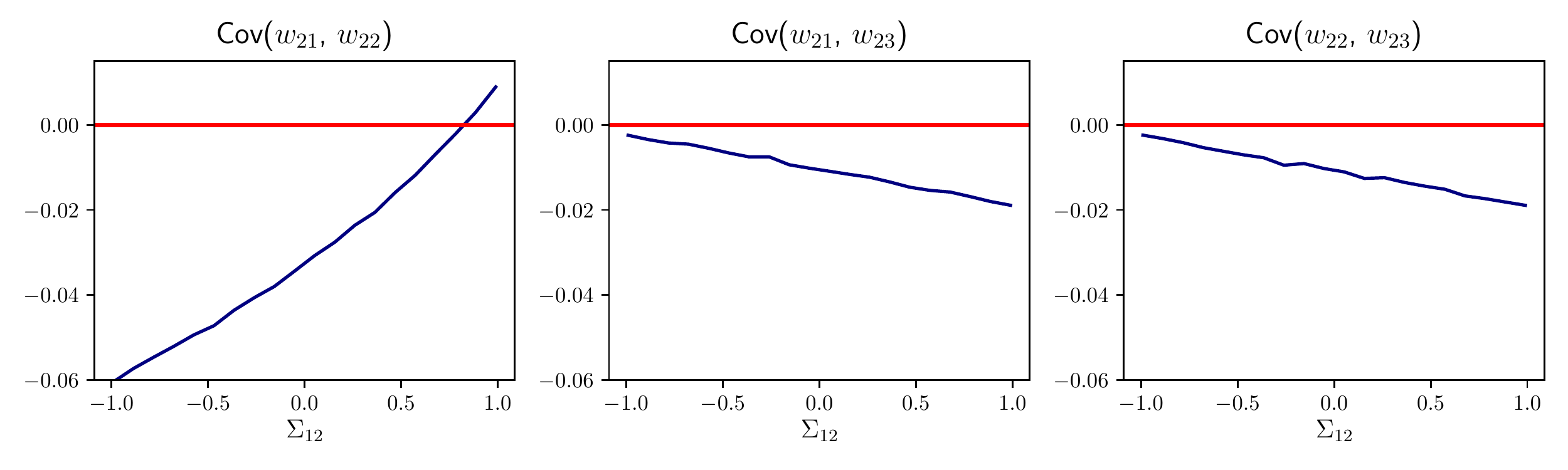}
    \caption{Pairwise covariance values of the components of $\bm w_2 = (w_{21}, w_{22}, w_{23})$ as a function of the correlation parameter $\Sigma_{12}$. The horizontal red line indicates the value 0.}
    \label{fig:cov_theta1}
\end{figure}

Figure~\ref{fig:cov_theta1} shows the covariance between the three components of $\bm w_2 = (w_{21}, w_{22}, w_{23})$ as a function of the correlation parameter $\Sigma_{12}$ in the matrix $\Sigma$ in \eqref{eq:joint_car}, having simulated $N=10,000$ MC draws.
Note that, unlike the finite-dimensional Dirichlet distribution,  the logistic-normal distribution  may have positive covariance among the components.
\begin{figure}[h!]
    \centering
    \includegraphics[width=\linewidth]{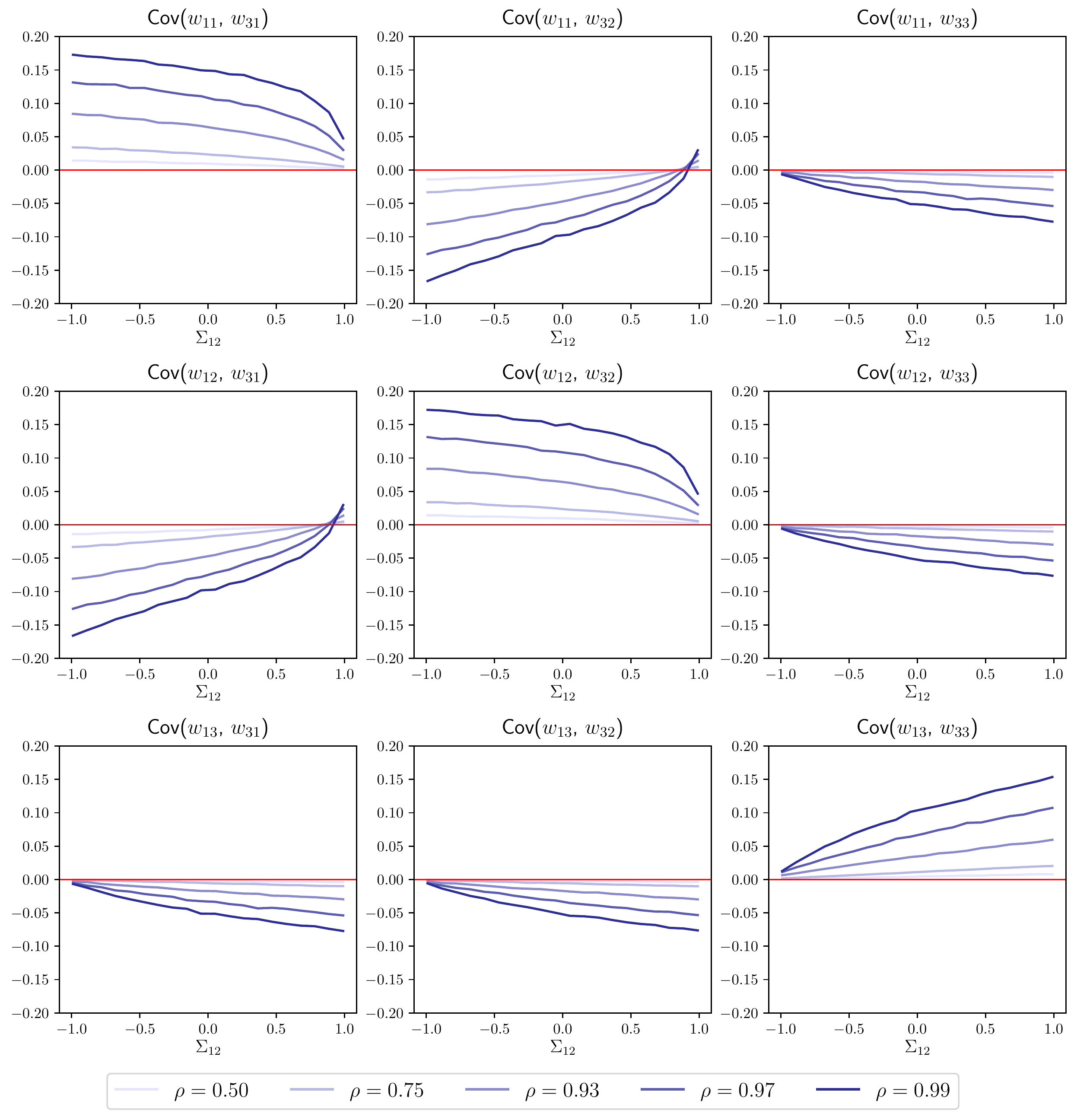}
    \caption{Pairwise  covariance values between components of $\bm w_1$ and $\bm w_3$, as a function of the correlation parameter $\Sigma_{12}$, for different values of $\rho$, when $(\bm w_1,\ldots, \bm w_5)$ has the logisticMCAR distribution. The horizontal red line indicates the value 0.}
    \label{fig:cov_theta1_theta3}
\end{figure}

Figure~\ref{fig:cov_theta1_theta3} instead shows the covariance between all the possible pairs $(w_{1j}, w_{3m})$ for $j,m=1, 2, 3$,  for different values of the parameter $\rho$.
%\bale Sull'asse delle $x$ c'\`{e} semple $\Sigma_{12}$ ma a linee diverse corrispondono valori di $\rho$ diversi?  \`E chiaro chela covarianza diventa positiva per certe combinazioni di $\rho$ e $\Sigma_{12}$ma non \`{e} mai fortissima. Vorrei vedere un grafico con $\Sigma_{12}$ fissato ad un valore molto alto, ma con $\rho$ sull'asse delle $x$) \eale
The covariances  between corresponding entries, i.e. $(w_{1j}, w_{2j}$ $j=1, 2, 3$) is
always positive, as expected since the spatial correlation parameter $\rho$ is 
always fixed to a positive value. 
The marginal prior for $\bm w_1, \bm w_3$ is exchangeable, since
nodes 1 and 3 belong to the same connected component in $G$.
This explains the symmetries in Figure~\ref{fig:cov_theta1_theta3}.
%\bale Ho tagliato una frase qui.\eale  
%On the other hand, it is clear that two vectors $\bm \wtilde_i$ and $\bm \wtilde_j$ are independent if they belong to different connected components of the graph $G$. 

In order to measure the association induced by our logisticMCAR prior, we  simulate the distances (Euclidean) of two vectors drawn from the joint distribution. 
\begin{figure}[h!]
    \centering
    \includegraphics[width=\linewidth]{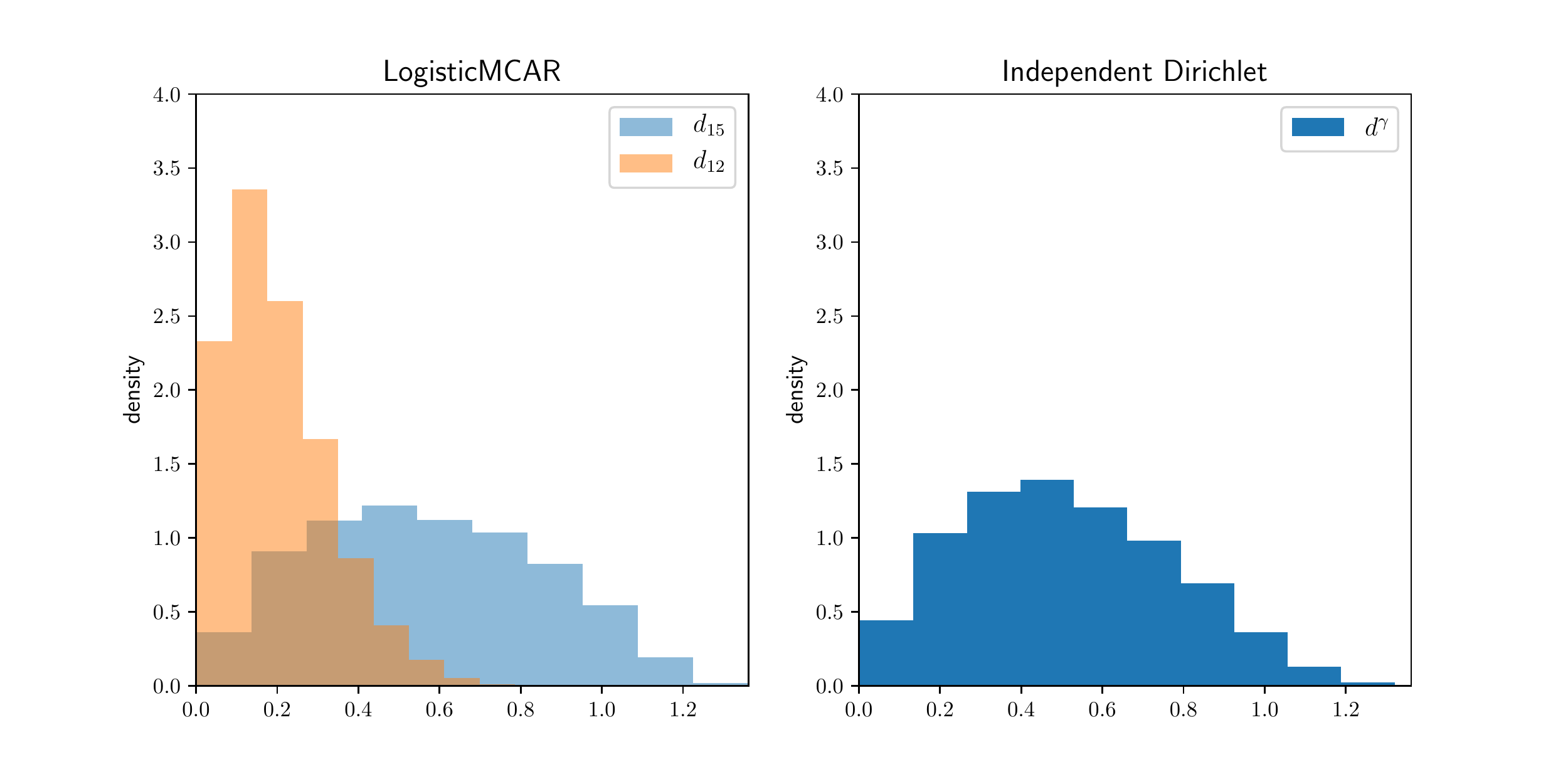}
    \caption{Histogram of MC draws from the marginal distributions of
  $d_{12}$ (orange) and $d_{15}$ (light blue) on the left and  from the marginal distribution of $d^{\gamma}$ on the right.}
    \label{fig:distances_hist}
\end{figure}
In particular, we simulated $N=10,000$ draws from the full joint logisticMCAR distribution of $(\bm w_1,\ldots, \bm w_5)$ with parameters as above, fixing $\Sigma_{12} = 0.5$, and computed the Euclidean  distances $d_{12}=||\bm w_1 - \bm w_2||$ and $d_{15}=||\bm w_1 - \bm w_5||$.
As $\bm w_1$ and $\bm w_2$ belong to the same connected graph component while $\bm w_5$ belongs to another component, we expect $\bm w_1$ and $\bm w_2$ to be more similar than $\bm w_1$ and $\bm w_5$ belonging to separate components. Hence the distance $d_{12}$ should be smaller than $d_{15}$. 
Moreover, for comparison, we also simulated $N=10,000$ draws from the joint distribution of two independent finite-dimensional Dirichlet random variables, i.e.
\[
    (\bm \gamma_{1}, \bm \gamma_{2})_i \iid \text{Dir}(\bm 1) \times \text{Dir}(\bm 1) \quad i=1, \ldots N 
\]
and  computed their Euclidean distance as well, that we denote by $d^\gamma$.
Figure~\ref{fig:distances_hist} reports the histograms of the marginal distributions of $d_{12}$, $d_{15}$ on the left and $d^\gamma$ on the right.
It is clear that $d_{12}$ is substantially smaller than $d_{15}$,  as expected. Moreover, by comparing $d_{15}$ and $d^\gamma$, we see that their marginal distributions are very similar. See also the summary statistics of these marginal distributions  in Table~\ref{tab:summary_dist}.  
\begin{table}
\centering
\begin{tabular}{c||c|c|c|c|c|}
    & $\min$ & $q_{0.25}$ & $q_{0.5}$ & $q_{0.75}$ & $\max$ \\
    \hline
    $d_{12}$ & $4 \times 10^{-4}$ & 0.10 & 0.18 & 0.27 &0.86 \\
    $d_{15}$ & 0.01 & 0.33 & 0.55 & 0.77 & 1.36 \\
    $d_\gamma$ & 0.007 & 0.31 & 0.52 & 0.71 & 1.36
\end{tabular}
\caption{Summary statistics of the marginal distributions of the distances $d_{12}, d_{15}, d_\gamma$, estimated from the MC
samples; $q_\alpha$ denotes the $\alpha$-quantile. 
\label{tab:summary_dist}
}
\end{table}

\begin{figure}[h!]
    \centering
    \includegraphics[width=\linewidth]{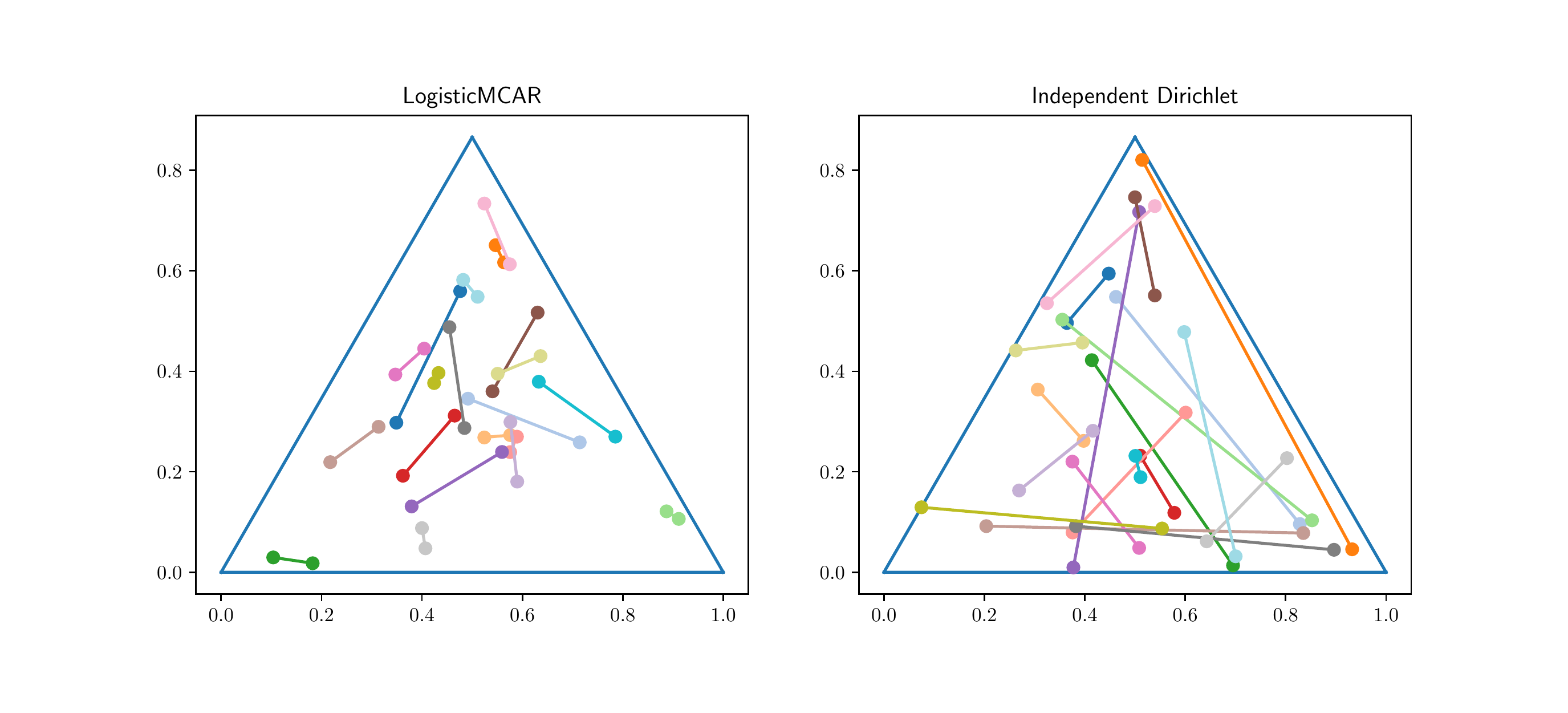}
    \caption{Plots of $N=20$ MC draws from the logisticMCAR distribution (left) and of $N=20$ MC draws from $\text{Dir}(\bm 1) \times \text{Dir}(\bm 1)$ (right). Each draws is represented with two dots (the values of the two random vectors) together with a colored line connecting them for visual purposes. Different colors correspond to independent draws.}
    \label{fig:triplot}
\end{figure}

For more insight, we report a subsample of size $N=20$ of the MC simulated values from the marginal distributions of $(\bm w_1, \bm w_2)$ and $(\bm \gamma_1, \bm \gamma_2)$, plotted on the  two dimensional projection of the simplex $S^3$ in Figure~\ref{fig:triplot}. Each pair is  denoted by two points inside the triangle and a line connecting them.
It is clear  that simulated values from $\Law(\bm w_1, \bm w_2)_i$ are much closer each other than those from $\Law(\bm \gamma_1, \bm \gamma_2)$.

\section{The Gibbs sampler}
\label{sec:app_Gibbs} \hfill\\ \noindent
\underline{\textbf{Proof of Equations \eqref{eq:full_cond_wih_pg} - \eqref{eq:full_cond_omegaih}}} \\
We start by writing the full conditional distribution \eqref{eq:full_cond_weights} as follows:
\begin{align*}
    \Law(\wtilde_{ih} \mid \bm \wtilde_{-i}, \bm \wtilde_{i, -h}, \bm s_i, \rho, \Sigma) & \propto \calN(\wtilde_{ih} \mid \mu^*_{ih}, \Sigma^*_{ih}) 
    \left( \frac{\e^{\eta_{ih}}}{1+\e^{\eta_{ih}}}\right)^{N_{ih}}
    \left( \frac{1}{1+\e^{\eta_{ih}}}\right)^{N_i - N_{ih}} \\
    & \propto \calN(\wtilde_{ih} \mid \mu^*_{ih}, \Sigma^*_{ih})  \frac{(\e^{\eta_{ih}})^{N_{ih}}}{(1+\e^{\eta_{ih}})^{N_i}} \\
    & \propto \calN(\wtilde_{ih} \mid \mu^*_{ih}, \Sigma^*_{ih}) \, \e^{(N_{ih} - N_i/2)\eta_{ih}}\int_0^\infty \e^{-\omega_{ih} \eta^2_{ih}/2} p(\omega_{ih}) d\omega_{ih}
\end{align*}
where $\omega_{ih} \sim PG(N_i, 0)$. We now include the latent variable $\omega_{ih}$ and derive the conditional distribution of $\wtilde_{ih}$, conditioning also to $\omega_{ih}$. We have
\begin{align*}
    \Law(\wtilde_{ih} \mid \bm \wtilde_{-i}, \bm \wtilde_{i, -h}, \bm s_i, \rho, \Sigma, \omega_{ih}) & \propto \calN(\wtilde_{ih} \mid \mu^*_{ih}, \Sigma^*_{ih})\, \e^{(N_{ih} - N_i/2)\eta_{ih}} \e^{-\omega_{ih} \eta^2_{ih}/2} \\
    & \propto \e^{-\frac{E}{2}}
\end{align*}
where 
\begin{align*}
    E &= \frac{(\wtilde_{ih}-\mu_{ih}^*)^2}{\Sigma^*_{ih}} - (2N_{ih}-N_i)(\wtilde_{ih}-C_{ih}) + \omega_{ih}(\wtilde_{ih}-C_{ih})^2 \\
    & \propto \wtilde_{ih}^2\left(\frac{1}{\Sigma^*_{ih}} + \omega_{ih} \right) -2 \wtilde_{ih} \left( \frac{\mu_{ih}^*}{\Sigma^*_{ih}} + N_{ih}-N_i/2 + \omega_{ih}C_{ih} \right) \\
    & \propto \left(\frac{1}{\Sigma^*_{ih}}+\omega_{ih}\right) \left(\wtilde_{ih}^2 -2 \wtilde_{ih} \left( \frac{\mu_{ih}^*}{\Sigma^*_{ih}} + N_{ih}-N_i/2 + \omega_{ih} C_{ih} \right) \left(\frac{1}{\Sigma^*_{ih}}+\omega_{ih}\right)^{-1}  \right)
\end{align*}
Thus
\begin{equation*}
    \Law(\wtilde_{ih} \mid \bm \wtilde_{-i}, \bm \wtilde_{i, -h}, \bm s_i, \rho, \Sigma, \omega_{ih}) \sim \calN(\hat\mu_{ih}, \hat \Sigma_{ih})
\end{equation*}
where
\begin{align*}
    \hat\mu_{ih} = \left(\frac{\mu_{ih}^*}{\Sigma^*_h} + N_{ih}-N_i/2 + \omega_{ih}C_{ih}\right)\left(\frac{1}{\Sigma^*_h}+\omega_{ih}\right)^{-1} 
    \quad
    \hat \Sigma_{ih} = \left(\frac{1}{\Sigma^*_h}+\omega_{ih}\right)^{-1}
\end{align*}

For the full conditional of $\omega_{ih}$ instead, it is sufficient to apply
Theorem 1 in \cite{polson2013bayesian} with $\psi = \eta_{ih}$ to obtain that
the law of $\omega_{ih}$, conditional to $\bm \wtilde_i$ is a P\'olya-Gamma distribution, i.e.
the density of $\omega_{ih}$ can be expressed as in Equation~\eqref{eq:pg_distrib},
with parameters $b=N_i$, $c=\wtilde_{ih} - \log \sum_{k \neq h} \exp(\wtilde_{ik})$.

\bigskip
\noindent
\underline{\textbf{Detailed description of the Gibbs sampler}} \\
The state of the MCMC sampler is made of $\bm \tau = (\tau_1, \ldots, \tau_H)$, $(\bm \wtilde_1, \ldots, \bm \wtilde_I)$, where $\bm \wtilde_i = \alr(\bm w_i)$, $\{s_{ij}\}_{ij}$ and $\bm \mtilde_{C_1}, \ldots \bm\mtilde_{C_k}$.
%We denote with $\bm x_{i,-h}$ the vector $\bm x_i$ except for the entry in position $h$.
The Gibbs sampler is obtained repeatedly sampling from the following conditional distributions:
\begin{itemize}
    \item For any $i=1,\ldots,I$ and $j=1,\ldots,N_i$, independently update the cluster allocation variables from
    \[
         p(s_{ij}=h \mid rest) \propto \text{alr}^{-1}(\wtilde_{ih}) \,k(y_{ij}\mid \tau_h) \quad h=1,\ldots,H
    \]
    \item Independently update the atoms of the mixture from
    \[
         \Law(\tau_h \mid rest) \propto P_0(\tau_h) \prod_{ij: s_{ij}=h}k(y_{ij}\mid \tau_h) \quad h=1,\ldots,H
    \]
    \item Sample $\Sigma$ from 
    \[
        \Law(\Sigma \mid rest) \propto \Law(\bm \wtilde \mid rest) \Law(\Sigma)
    \]
    We show that the full conditional of $\Sigma$ is still an inverse-Wishart distribution.
    To see this, write the right hand side as follows
    \begin{align*}
        \Law(\Sigma \mid rest) & \propto |(F-\rho G)^{-1} \otimes \Sigma|^{-1/2} \exp \left(-\frac{1}{2} (\bm \wtilde - \bm \mtilde)^T \left( (F- \rho G) \otimes \Sigma^{-1} \right) (\bm \wtilde - \bm \mtilde) \right)  \\
        & \qquad \times |\Sigma|^{-(\nu + (H-1) + 1) / 2} \exp \left(-\frac{1}{2} tr(V \Sigma^{-1}) \right)
    \end{align*}
    Now $|(F-\rho G)^{-1} \otimes \Sigma| = |(F - \rho G)^{-1}|^{H-1} \times |\Sigma|^{I}$, so that the degrees of freedom in the full conditional are  $\nu_p = \nu + I$.
    Working on the exponent, write the quadratic form involving the Kronecker product as follows
    \[
        (\bm \wtilde - \bm \mtilde)^T \left( (F- \rho G) \otimes \Sigma^{-1} \right) (\bm \wtilde - \bm \mtilde)  = \sum_{i, j = 1}^{I} (F - \rho G)_{ij} (\bm \wtilde_i - \bm \mtilde_i)^T \Sigma^{-1} (\bm \wtilde_j - \bm \mtilde_j)
    \]
By exploiting multiple times the linearity of the trace operator and its cyclic property, the scale matrix $V_p$ can be seen to equal
    \[
        V_p = \sum_{i, j = 1}^I (F - \rho G)_{ij} (\bm \wtilde_j - \bm \mtilde_j) (\bm \wtilde_i - \bm \mtilde_i)^T + V
    \]
    and we conclude that $\Sigma \mid rest \sim \text{Inv-Wishart}(\nu_p, V_p)$ 
    \item Sample $\rho$ from its full conditional:
    \[
        \Law(\rho \mid rest) \propto \pi(\rho) \calN(vec(\wtilde_1, \dots, \wtilde_I) \mid \bm 0, (F - \rho G)^{-1} \otimes \Sigma)
    \]
This distribution does not have a closed form analytic expression because the support of $\rho$ is $(0,1)$ and hence we resort to a Metropolis Hastings step.
    The proposal distribution is a truncated normal (with support on $(0,1)$) centered in the current value of $\rho$ with standard deviation $0.1$.
    Sampling from the truncated normal is performed by rejection sampling, whereas the computation of the acceptance rate for the Metropolis Hastings step is obtained by exploiting the law of the matrix normal distribution, which does not require to factorize the matrix $(F - \rho G)^{-1} \otimes \Sigma$. 
To improve the mixing of the chain,  we resort to 
    an Adaptive Metropolis Hastings move as in \cite{roberts2009examples} to automatically tune  variance of the 
    normal proposal distribution.

%    We are pretty satisfied with the results produced, see Figure~\ref{fig:rhochain}
%    for the example of one traceplot.
%    }
    \item For each $i = 1, \ldots, I$ and each $h = 1, \ldots H$, independently sample $\wtilde_{ih}$ as follows:
    \begin{itemize}
        \item Sample the latent variable $\omega_{ih}$ from 
        \[
         \Law(\omega_{ih} \mid \bm \wtilde_i) = PG(N_i, \eta_{ih}) = PG\left(N_i, \wtilde_{ih}- \log \sum_{k \neq h}e^{\wtilde_{ik}}\right)
        \]
        \item Sample the transformed weight $\wtilde_{ih}$ from
        \[
        \Law(\wtilde_{ih} \mid \bm \wtilde_{-i}, \bm \wtilde_{i, -h}, \bm s_i, \rho, \Sigma, \omega_{ih}) = N(\hat\mu_{ih}, \hat \Sigma_{ih}).
        \]
    \end{itemize}
    
\item for each connected component $m$ of the graph we sample from   
		\[
		\Law(\bm\mtilde_{C_m}\mid rest) = \mathcal{N} (\bm m_{C_m},\Lambda_{C_m})
		\]
For ease of notation, we show how to obtain expression of $\bm m_{C_m}$ and $\Lambda_{C_m}$ in the case where is only one connected component in the graph. However the general update can be straightforwardly recovered since 
   $\mtilde_{C_1}, \ldots \bm\mtilde_{C_k}$ corresponding to connected components in the graph  are conditionally independent a priori. 
%% ones just need to replace the indexes and the prior on the weights, with the ones regarding only the selected connected component.
In case of one single connected component in the graph,  we rewrite \eqref{eq:joint_car}, letting all the $\bm \mtilde_i$s to be equal to $\bm \mtilde_1$, as 
\[
    \bm \wtilde \sim \calN_{I (H-1)} \left( {\mathbf{1}}_I \otimes \mathbb{I}_{H-1} \bm \mtilde_1, \left((F - \rho G) \otimes \Sigma^{-1} \right)^{-1} \right)
\]
where $\mathbf{1}_I$ is the vector of ones of length $I$ and ${\mathbb{I}}_{H-1}$ is the $(H-1) \times (H-1)$
identity matrix.
Then if $\Lambda: = diag(\sigma^2, \ldots, \sigma^2)$ and writing ${\mathbb{I}}^* =  \mathbf{1}_I \otimes I_{H-1}$, $Q = (F - \rho G) \otimes \Sigma^{-1}$, we can write the full conditional of
$\bm \mtilde_1$ as follows:
\begin{align*}
    \Law(\bm \mtilde_1 \mid rest) \propto & \exp \bigg( -0.5 \big(\bm \wtilde - {\mathbb{I}}^* \bm \mtilde_1)^T Q (\bm \wtilde -{\mathbb{I}}^* \bm \mtilde_1)^T  + \bm \mtilde_1^T \Lambda^{-1} \bm \mtilde_1 \big) \bigg) \\
    \propto & \exp \bigg(-0.5 \bm \mtilde_1^T \left( {\mathbb{I}}^{*T} Q {\mathbb{I}}^* \right) \bm \mtilde + \bm \mtilde_1^T \Lambda^{-1} \bm \mtilde_1 +  
     -2 \bm \mtilde_1^T \left({\mathbb{I}}^{*T} Q \bm \wtilde \right)\bigg)
\end{align*}
This is the kernel of a multivariate normal distribution with 
covariance matrix $\Lambda_C = \left( {\mathbb{I}}^{*T} Q {\mathbb{I}}^* + \Lambda^{-1} \right)^{-1}$ and mean $\bm m_C = \Lambda_C \left({\mathbb{I}}^{*T} Q \bm \wtilde \right)$.

\item If there are covariates in the model$ M1$ as in Section~\ref{sec:airbnb}, the full-conditional of the regression coefficients $\bm \beta$ is given by
\[
    \Law(\bm \beta \mid rest) = \calN_d\left((\Sigma^{-1} + X^T V X)^{-1}(X^TV(\bm y -\bm \mu)) ,(\Sigma^{-1} + X^T V X)^{-1}\right)
\]
where $V$ is an $N\times N$ diagonal matrix and $N=\sum N_i$. Denoting by $\bm c = (c_1, \ldots, c_N)$, the vectorization 
of the sequence of latent vectors $\bm s_1, \ldots, \bm s_I$ in \eqref{eq:lik_latent}-\eqref{eq:cluster_allocs}, then
one has $V_{k, k} = \sigma^2_{c_{k}}$.
 The formula above can be derived by standard posterior updates in the Bayesian linear regression,  when the mixture model \eqref{eq:lik}-\eqref{eq:prior_rho}  is the model for the \virgolette{regression error}.
 
In case of model $M2$,  as in Section~\ref{sec:airbnb}, the full-conditional of each regression coefficients   $\left((\mu_h, \bm \beta_h), \sigma^2_h\right)$ is straightforwardly computed from standard Bayesian linear regression, considering only observations that are allocated to component $h$.
In particular, if we denote by $\bm y_h$ the vector of $\{y_{ij}: s_{ij} = h\}$, by $X_h$ the matrix with rows $\{x_{ij}: s_{ij} = h\}$, and by $n_h$ the size of $\bm y_h$, then we have
\begin{equation*}
  \begin{aligned}
    \sigma^2_h \mid rest &\sim IG(a_{ph}, b_{ph}) \\
    (\mu_h, \beta_h) \mid \sigma^2_h, rest & \sim \mathcal{N}(\bm \mu_{ph}, \Delta_{ph})
  \end{aligned}
\end{equation*}
where
\begin{equation*}
  \begin{aligned}
    \Delta_{ph} &= X_{h}^T X_{h} + 10 \mathbb{I}_{d+1} \\
    \bm \mu_{ph} &= \Delta_{ph}^{-1} X_h^T \bm y_{h} \\
    a_{ph} &= 2 + n_h / 2  \\
    b_{ph} &= 2 + \frac{1}{2}(\bm y_h^T \bm y_h - \bm \mu_{ph}^T \Delta_{ph} \bm \mu_{ph} )
  \end{aligned}
\end{equation*}

\end{itemize}

\section{Additional plots and tables}
\label{sec:app_additionalplots}

\begin{figure}[h!]
    \centering
    \includegraphics[width=\linewidth]{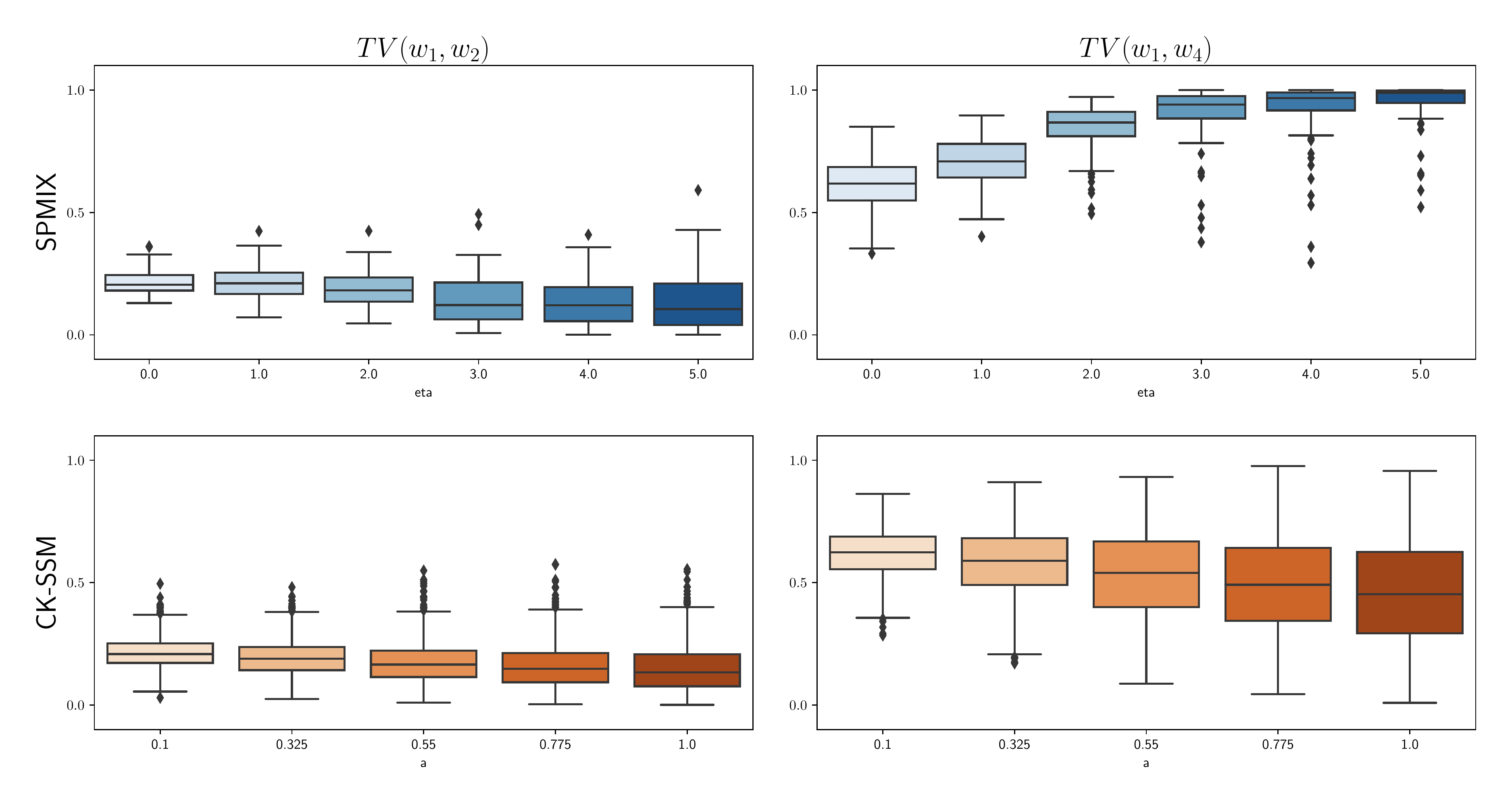}
    \caption{Total variation distances between the vectors $(\bm w_1, \bm w_2)$ and $(\bm w_1, \bm w_4)$ under the logisticMCAR distribution (first row) and the CK-SSM(second row). Each plot shows the boxplots of 1,000 independents simulations, for different values of the sparsity-tuning parameters (sparsity is increasing from left to right in each plot). The remaining hyperparameters and the adjacency matrix are as discussed in Section~\ref{sec:comparison}.}
    \label{fig:prior_dist}
\end{figure}

\begin{figure}[h!]
    \centering
    \includegraphics[width=\linewidth]{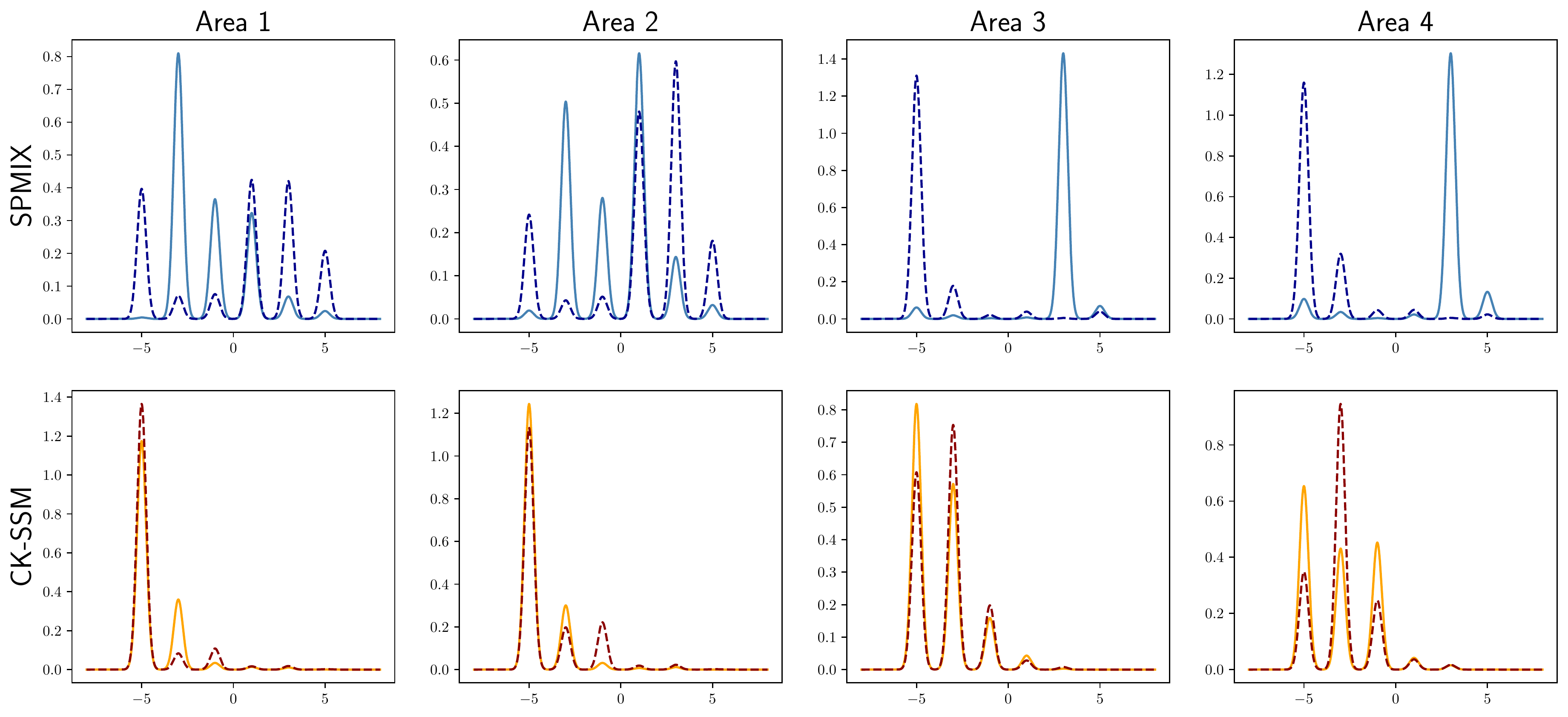}
    \caption{Samples from the prior distribution with $H=6$  under the logisticMCAR (first row, with $\eta=5.0$) and CK-SSM in \cite{jo2017dependent} (second row, with $b=0.5, a = 1.0$). The remaining hyperparameters and the adjacency matrix are as discussed in Section~\ref{sec:comparison}. Each plot shows the mixture density in one particular area and different line types / colors represent independent draws from the prior. Here the means of the mixture model are fixed as $\mu_1 = -5$, $\mu_2 = -3.33, \ldots, \mu_6 = 5$ and the variances are all equal to $0.25^2$.}
    \label{fig:prior_dens}
\end{figure}

\begin{figure}[h!]
	\centering
	\includegraphics[width=0.5\textwidth]{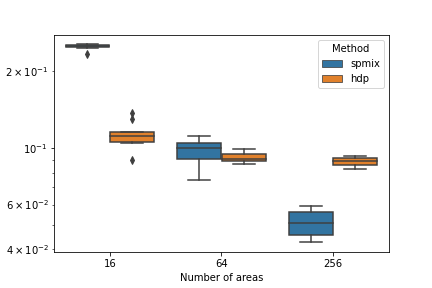}
	\caption{Boxplots of the Hellinger distance between true density \eqref{eq:spatial_truedens} and estimated one under our model (spmix) and the HDP-mixture model (hdp) for each simulation, averaged over the areas, for $I=16, 64, 256$, , in logarithmic scale, in Section~\ref{sec:spatial_weights}.}
	\label{fig:error_sim2_Hellinger}
\end{figure}

\begin{figure}[h!]
	\centering
	\includegraphics[width=0.5\linewidth]{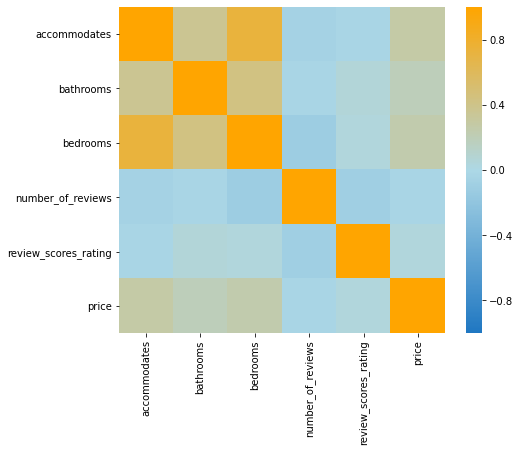}
	%\vspace*{-1mm}
	\caption{Correlation matrix between numerical predictors and response for Airbnb Amsterdam}
	\label{fig:corr_matrix}
\end{figure}

\begin{figure}[h!]
	\centering
	\includegraphics[width=\linewidth]{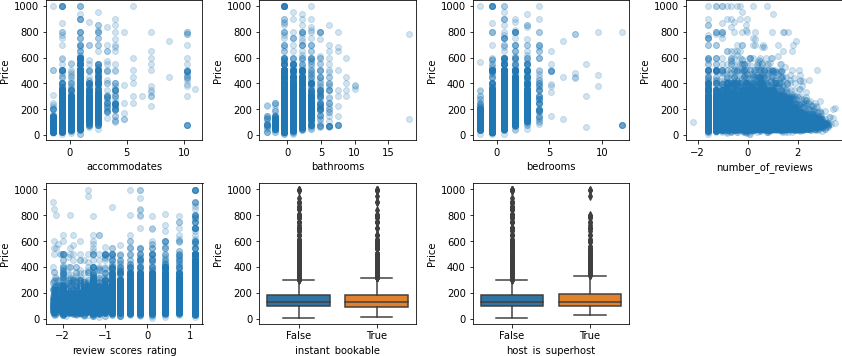}
	%\vspace*{-1mm}
	\caption{Scatterplots and boxplots of the nightly price versus predictors for Airbnb Amsterdam.  Numerical predictors have been standardized.}
	\label{fig:predictors_plots}
\end{figure}

\begin{figure}[h!]
\centering
   \includegraphics[width=0.4\linewidth]{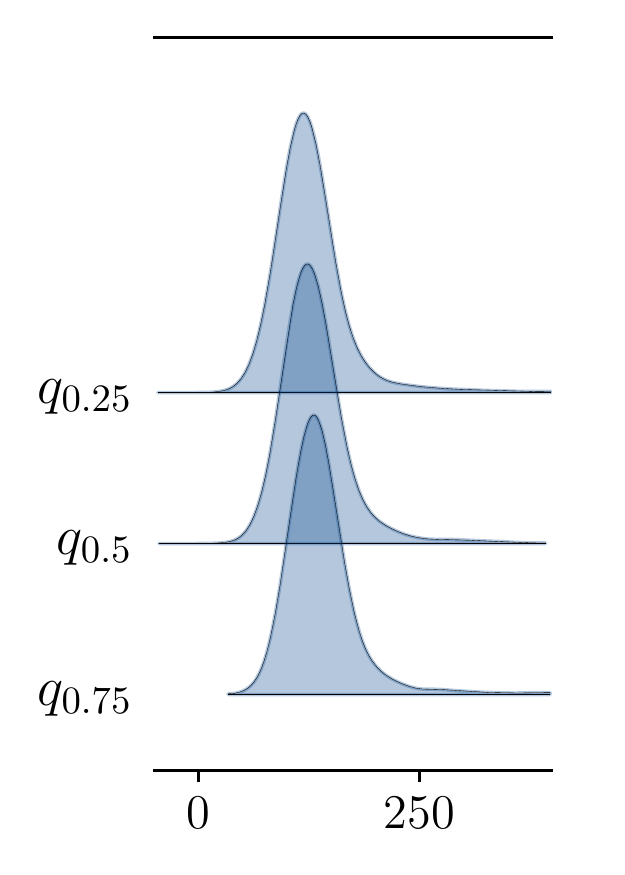}
   \caption{Predictive density for a new listing in \emph{Bijlmer-Centrum} with all numerical covariates fixed
        to the empirical median of the dataset except \texttt{reviews\_scores\_rating}
         that ranges in the values $q_{0.25}, q_{0.5}, q_{0.75}$, where $q_\alpha$
         denotes the empirical quantile of order $\alpha$.
        Each line corresponds to one of these values, from top to bottom.}
   \label{fig:bijlmer}
\end{figure}

\begin{itemize}
\item Figure~\ref{fig:prior_dist} shows the total variation distance for $(\bm w_1, \bm w_2)$ and  $(\bm w_1, \bm w_4)$ under the logisticMCAR and the prior in \cite{jo2017dependent} with parameters as in Section~\ref{sec:comparison}. Observe how the distance between $(\bm w_1, \bm w_2)$ decreases as the sparsity increases under both priors. This is expected since areas 1 and 2 are neighbors. However, the distance between $(\bm w_1, \bm w_4)$ increases with sparsity under the logisticMCAR but decreases under CK-SSM.  Hence, under CK-SSM, forcing sparsity in the mixture model results in imposing similar behaviors to different connected components.

\item Figure~\ref{fig:prior_dens} shows draws from the prior mixture density corresponding to parameters sampled from the prior under the logisticMCAR and CK-SSM, having fixed the atoms to have means $\mu_1 = -5, \ \mu_2 = -3.33, \ldots, \mu_6=5$ and equal variances $0.25^2$ and remaining hyperparameters as in Section~\ref{sec:comparison}. It is clear that the logisticMCAR prior allows great variety among disconnected components as well as across different independent samples. Instead, CK-SSM shows that only the first 2/3 components have a nonzero weight, so that the densities across different areas and coming from independent samples are also similar.

\item Table~\ref{tab:results_hell}
shows the Hellinger distance  between the true density and the estimate under the three models under comparison in Section~\ref{sec:nongaussian} for the three simulated scenarios in Table~\ref{tab:sim_01_data} for 100 repeatedly simulated datasets. We average these values over the simulated datasets, also considering $\pm$ one empirical standard deviation of the 100 values obtained.

\item Figure~\ref{fig:error_sim2_Hellinger} shows errors, measured with the Hellinger distance, under our model (spmix) and the HDP-mixture model (hdp) for each simulation, averaged over the areas, for $I=4, 64, 256$, in Section~\ref{sec:spatial_weights}.

\item Figure~\ref{fig:corr_matrix} displays empirical correlations  among the predictors and, in the last column, between predictors and the response for the Airbnb Amsterdam dataset in Section~\ref{sec:airbnb}.

\item Figure~\ref{fig:predictors_plots} shows the scatterplots of the response price versus numerical predictors and boxplots for categorical predictors for the Airbnb Amsterdam dataset in Section~\ref{sec:airbnb}.

\item Figure~\ref{fig:bijlmer} shows the predictive densities in area Bijlmer-Centrum, corresponding to different covariate specifications: 
all the covariates are fixed  to their empirical median
except for \texttt{reviews\_scores\_rating},  which assumes values equal to the  empirical quartiles $q_{0.05}, q_{0.5}, q_{0.95}$.
It is clear that the three densities overlap almost perfectly.
There are two reasons for this. First, the the empirical distribution of this covariate  is 
it is highly concentrated around high values, as people tend to give mostly positive
reviews. 
Second, the coefficient associated to this covariate, despite significant, has
a very small absolute value.

\item Table~\ref{tab:p200} shows the posterior predictive probability $P(y^\star > 200 \mid x^\star, i)$ for the same neighborhoods and values of $x^\star$ considered in Figure~\ref{fig:res_dens} (c).

\end{itemize}

\begin{table}[h!]
    \centering
    \resizebox{\textwidth}{!}{%
    \begin{tabular}{c|c|c|c|c|c|c|c}
          & Model & 1 &  2 &  3 &  4 &  5 &  6 \\
         \hline
         Scenario I & SPMIX & $0.06 \pm 0.01$ & $0.06 \pm 0.01$ & $0.06 \pm 0.01$ & $0.06 \pm 0.01$ & $0.09 \pm 0.01$ & $0.09 \pm 0.01$\\
& HDP & $0.03 \pm 0.01$ & $0.03 \pm 0.01$ & $0.06 \pm 0.01$ & $0.06 \pm 0.01$ & $0.09 \pm 0.01$ & $0.09 \pm 0.01$\\
& CK-SSM & $0.44 \pm 0.06$ & $0.44 \pm 0.06$ & $0.53 \pm 0.03$ & $0.53 \pm 0.03$ & $0.44 \pm 0.03$ & $0.44 \pm 0.03$\\
\hline
Scenario II & SPMIX & $0.08 \pm 0.01$ & $0.11 \pm 0.02$ & $0.07 \pm 0.01$ & $0.08 \pm 0.03$ & $0.11 \pm 0.00$ & $0.11 \pm 0.03$\\
& HDP & $0.04 \pm 0.01$ & $0.19 \pm 0.02$ & $0.09 \pm 0.01$ & $0.24 \pm 0.03$ & $0.10 \pm 0.00$ & $0.27 \pm 0.03$\\
& CK-SSM & $0.44 \pm 0.06$ & $0.43 \pm 0.06$ & $0.53 \pm 0.03$ & $0.53 \pm 0.03$ & $0.45 \pm 0.05$ & $0.45 \pm 0.05$\\
\hline
Scenario III & SPMIX & $0.20 \pm 0.07$ & $0.20 \pm 0.07$ & $0.16 \pm 0.06$ & $0.16 \pm 0.06$ & $0.11 \pm 0.05$ & $0.11 \pm 0.05$\\
& HDP & $0.12 \pm 0.07$ & $0.12 \pm 0.07$ & $0.21 \pm 0.06$ & $0.21 \pm 0.06$ & $0.13 \pm 0.05$ & $0.13 \pm 0.05$\\
& CK-SSM & $0.42 \pm 0.06$ & $0.42 \pm 0.06$ & $0.59 \pm 0.03$ & $0.59 \pm 0.03$ & $0.38 \pm 0.07$ & $0.38 \pm 0.07$\\ 
    \end{tabular}}
    \caption{Hellinger distances between the true densities and the estimated ones, aggregated over $100$ simulated datasets with $\pm$ one standard deviation for the simulated data in Section~\ref{sec:nongaussian}}
    \label{tab:results_hell}
\end{table}

\begin{table}
\begin{tabular}{c|c|c|c|c}
 & \vtop{\hbox{\strut Bijlmer}\hbox{\strut Centrum}} & 
\vtop{\hbox{\strut Gaasperdam}\hbox{\strut Driemond}} &
\vtop{\hbox{\strut Oostelijk Havengebied}\hbox{\strut Indische Buurt}} & \multirow{2}{*}{Watergraafsmeer} \\ 
\hline
$q_{0.05}$ & 0.02 & 0.01 & 0.05 & 0.05 \\
$q_{0.50}$ & 0.02 & 0.01 & 0.06 & 0.06 \\
$q_{0.95}$ & 0.06 & 0.05 & 0.26 & 0.27\\
\end{tabular}
\caption{Posterior posterior predictive probability $P(y_{ij} > 200)$ for the same neighborhoods and covariate choices as in Figure~\ref{fig:res_dens}.}
\label{tab:p200}
\end{table}

\FloatBarrier
%
%\end{supplement}

\bibliographystyle{ba}
\newlength{\bibitemsep}\setlength{\bibitemsep}{.05\baselineskip}
\newlength{\bibparskip}\setlength{\bibparskip}{0pt}
\let\oldthebibliography\thebibliography
\renewcommand\thebibliography[1]{%
  \oldthebibliography{#1}%
  \setlength{\parskip}{\bibitemsep}%
  \setlength{\itemsep}{0em}%
}
\renewcommand{\baselinestretch}{0.0}
\bibliography{references}  

\end{document}